\documentclass[%
 reprint,
superscriptaddress,
 amsmath,amssymb,
 aps,
pra,
]{revtex4-1}

\usepackage{graphicx}% Include figure files
\usepackage{epstopdf}
\usepackage{xcolor}
\usepackage{dcolumn}% Align table columns on decimal point
\usepackage{bm}% bold math
\usepackage{hyperref}% add hypertext capabilities
\hypersetup{colorlinks,linkcolor=blue,citecolor=blue}
\usepackage[''option'']{SIunits}

\usepackage[colorinlistoftodos]{todonotes}

\newcommand{\crad}{\ensuremath{\delta\left<r_c^2\right>^{AA'}} }%
\newcommand{\ba}{{\rm Ba}}

\newcommand{\hz}{\,{\rm Hz}}
\newcommand{\khz}{\,{\rm kHz}}

\newcommand{\ghz}{\,{\rm GHz}}
\newcommand{\amu}{\,{\rm amu}}
\newcommand{\ev}{\,{\rm eV}}

\begin{document}

\preprint{APS/123-QED, EFI-20-6, FERMILAB-PUB-20-142-T}

\title{Improved isotope-shift-based bounds on bosons beyond the Standard Model through measurements of the $^2$D$_{3/2} - ^2$D$_{5/2}$ interval in Ca$^+$}

\author{Cyrille Solaro}
\email{solaro@phys.au.dk}
\affiliation{Department of Physics and Astronomy, Aarhus University, DK-8000 Aarhus C, Denmark}
\author{Steffen Meyer}
\email{steffen.meyer@phys.au.dk}
\affiliation{Department of Physics and Astronomy, Aarhus University, DK-8000 Aarhus C, Denmark}
\author{Karin Fisher}
\affiliation{Department of Physics and Astronomy, Aarhus University, DK-8000 Aarhus C, Denmark}
\author{Julian C. Berengut}
\email{julian.berengut@unsw.edu.au}
\affiliation{School of Physics, University of New South Wales, Sydney NSW 2052, Australia}
\author{Elina Fuchs}
\email{elinafuchs@uchicago.edu}
\affiliation{Fermilab, Theory Department, Batavia, IL 60510, USA}
\affiliation{University of Chicago, Department of Physics, Chicago, IL 60637, USA}
\author{Michael Drewsen}
\email{drewsen@phys.au.dk}
\affiliation{Department of Physics and Astronomy, Aarhus University, DK-8000 Aarhus C, Denmark}

\date{\today}% It is always \today, today,

\begin{abstract}
We perform high-resolution spectroscopy of the $3$d$~^2$D$_{3/2} - 3$d$~^2$D$_{5/2}$ interval in all stable even isotopes of $^A$Ca$^+$ (A = 40,\,42,\,44,\,46\,and 48) with an accuracy of $\sim$ 20 Hz using direct frequency-comb Raman spectroscopy. Combining these data with isotope shift measurements of the 4s$~^2$S$_{1/2} \leftrightarrow 3$d$~^2$D$_{5/2}$ transition, we carry out a King plot analysis with unprecedented sensitivity to coupling between electrons and neutrons by bosons beyond the Standard Model. Furthermore, we estimate the sensitivity to such bosons from equivalent spectroscopy in Ba$^+$ and Yb$^+$. Finally, the data yield isotope shifts of the 4s$~^2$S$_{1/2} \leftrightarrow 3$d$~^2$D$_{3/2}$ transition at 10 part-per-billion through combination with recent data of Knollmann \textit{et al.} \cite{Knollmann2019}.
\end{abstract}

\maketitle

The Standard Model of particle physics (SM) cannot be complete 
since, e.g., it lacks a Dark Matter candidate, cannot produce the observed matter-antimatter asymmetry of the universe and does not explain the hierarchy between the Higgs mass and the Planck scale. Because the masses of new particles are unknown, searches for New Physics (NP) beyond the SM involve multiple frontiers (see e.g. Refs.~\cite{Mangano:2020icy,Bettoni:2006za,DeMille:2017azs} and references therein) ranging from high-energy colliders, high-intensity beam dumps, astrophysical and cosmological observations to high-precision table-top experiments. In the search for new long-range interactions, high-resolution spectroscopy of atoms and molecules is a driving force \cite{Safronova2018}. A recent example is to probe the existence of new bosons that couple to both nucleons and electrons from precisely measured isotope shifts. Conversely, agreement between the prediction based on the SM and experiments within their uncertainties allows for placing bounds on the coupling strength of the potential new interaction depending on the mass of the new boson. Except for few-electron systems \cite{Delaunay2017b}, the main limitation in translating the experimental accuracy to a stringent bound is the theory uncertainty. To mitigate this problem, 
Delaunay \textit{et al.} \cite{Delaunay2017a}  proposed to measure isotope shifts of two different transitions of the same element and to look for a non-linearity of the so-called King Plot \cite{King1963}. This allows to place bounds on long-range mediators \cite{Berengut2018}, and thus to test various particle physics models \cite{Frugiuele2017}. %independently of theory uncertainties. 
For instance, the protophobic model \cite{Feng:2016jff,Feng:2016ysn} of a new boson at 17\,MeV/$c^2$ for the Be anomaly \cite{Krasznahorkay:2015iga} is in reach of near-future Sr/Sr$^+$ and Yb$^+$ King plot analyses \cite{Frugiuele2017,Berengut2018}. 
This data-driven method requires only theory input for the new interaction, but is independent of SM multi-electron and nuclear calculations - unless a non-linearity from higher-order SM effects is predicted at the level of experimental precision. After subtracting the predicted SM non-linearity, the residual non-linearity can be used to constrain a NP contribution.
A King plot, however, requires at least four isotopes (preferably with zero nuclear spin) in order to test the linearity of the isotope shifts of the resulting three independent isotope pairs. Calcium is in this respect a good candidate with the five stable, spin-0 isotopes A = 40, 42, 44, 46 and 48. Previously, Gebert \textit{et al.} reported measurements of two dipole allowed transitions, 
4s$~^2$S$_{1/2} \leftrightarrow 4$p$~^2$P$_{1/2}$ (397-nm) and 
3d$~^2$D$_{3/2} \leftrightarrow 4$p$~^2$P$_{1/2}$ (866-nm), 
in the four $^{40,42,44,48}$Ca$^+$ isotopes with an accuracy of $\mathcal{O}(100)$ kHz corresponding to a fractional accuracy on the isotope shifts in the $10^{-5}-10^{-4}$ range \cite{Gebert2015}. In principle, far better accuracy can be achieved on narrow-optical transitions \cite{Manovitz2019} such as the two 4s - 3d quadrupole transitions. While the 
4s$~^2$S$_{1/2} \leftrightarrow 3$d$~^2$D$_{5/2}$ (729-nm) 
transition has been measured at the Hz level in $^{40}$Ca$^+$ \cite{Chwalla2009}, measurement of the
4s$~^2$S$_{1/2} \leftrightarrow 3$d$~^2$D$_{3/2}$ (732-nm) transition 
at the same level is more challenging since the electron-shelving technique \cite{Dehmelt1982} cannot directly be used for state detection.

In this letter, we report isotope shift measurements of the 
$3$d$~^2$D$_{3/2} - 3$d$~^2$D$_{5/2}$ interval (i.e. the D-fine-structure splitting isotope shift (DSIS)) on all five stable even isotopes of $^A$Ca$^+$ (with A = 40 as the reference isotope) using direct frequency-comb Raman spectroscopy \cite{Solaro2018}. Combining these with isotope shift measurements of the 4s$~^2$S$_{1/2} \leftrightarrow 3$d$~^2$D$_{5/2}$ transition (729-IS), we deduce the isotope shift of the 4s$~^2$S$_{1/2} \leftrightarrow 3$d$~^2$D$_{3/2}$ transition (732-IS). 
This leads to a King plot analysis with unprecedented sensitivity to NP bosons coupling to both electrons and neutrons. In addition, the analysis yields a field shift ratio of the 729-nm and 732-nm transitions with an unprecedented fractional accuracy of $2\times 10^{-7}$. We achieve an absolute accuracy on the DSIS at the 20 Hz level corresponding to a fractional accuracy in the $10^{-6}$ range, and on the 729-IS at the 2 kHz level corresponding to a fractional accuracy in the $10^{-7}$ range. 
We show that, with respect to bounds on NP bosons, our measurements are in fact equally precise as measuring the isotope shift of the two 4s - 3d transitions with the same 20 Hz level accuracy, since the limiting fractional accuracy is the DSIS measurement. In particular, the King plot analysis is not improved through combination with recent 729-IS measurements at the $10^{-9}$ level by Knollmann \textit{et al.} \cite{Knollmann2019}. It is neither limited by the 729-IS involving the isotope $^{46}\text{Ca}^+$ that was not measured in Ref.~\cite{Knollmann2019}. However, the combined data yield isotope shifts of the 732-nm transition with fractional accuracy below the $10^{-8}$ level.

\begin{figure}
	\centering
	\includegraphics[width=.48\textwidth]{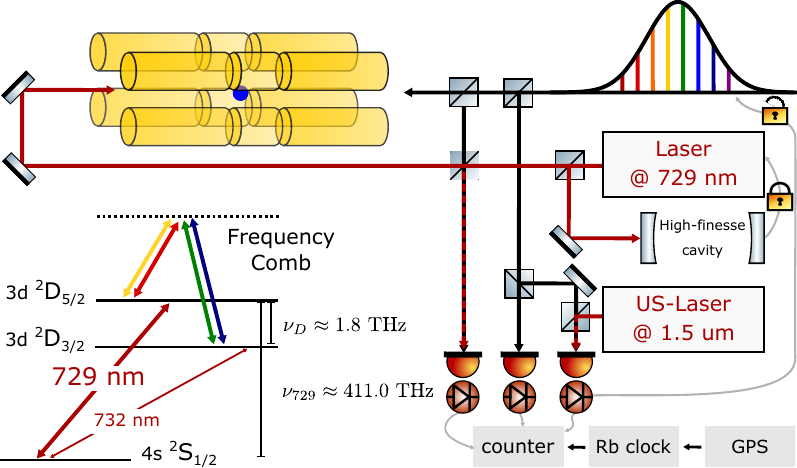}
	\caption{Schematic of the experimental setup and the relevant electronic levels of the Ca$^+$ ion. The isotope shift of the 732-nm transition can be deduced from the isotope shifts of the 729-nm transition and of the D-fine-structure splitting. The D-fine-structure splitting is measured successively on the different calcium isotopes by direct frequency-comb Raman spectroscopy \cite{Solaro2018}. The transition frequency is deduced from the measurement of the comb repetition rate on a frequency counter referenced to a GPS-disciplined rubidium standard. The 729-nm laser used to probe the 4s$~^2$S$_{1/2} \leftrightarrow 3$d$~^2$D$_{5/2}$ transition is locked to an ultra-stable high-finesse cavity providing a short-term linewidth $<1$ kHz. The absolute laser frequency is deduced from a measurement on the frequency counter of the beating between the laser and the frequency comb with the latter locked to an ultra-stable laser at 1.5 $\micro$m.}
\label{fig:setup}
\end{figure}

The splitting isotope shift of the 3d$~^2$D$_{3/2}$ and 3d$~^2$D$_{5/2}$ states $\delta\nu^{A,40}_\text{DSIS}$ was measured using direct frequency-comb Raman spectroscopy, as described in details in \cite{Solaro2018}. In brief, a single Ca$^+$ isotope is loaded into a linear Paul trap via isotope-selective photoionization in a neutral calcium beam \cite{Kjergaard2000,Mortensen2004}. An external magnetic field of 6.500(3) G lifts the Zeeman degeneracy of the involved electronic energy levels by a few MHz, allowing for Zeeman-resolved spectroscopy of the D$_{3/2}$-D$_{5/2}$ interval. The experimental cycle is initialized by Doppler cooling, followed by sideband cooling and finally optical pumping of the Ca$^+$ ion into one of its $|4s~^2S_{1/2},m_j=\pm1/2\rangle$ states. Next, the ion is prepared to the $|D_{5/2}, m_j=\pm 1/2\rangle$ state using rapid adiabatic passage (RAP) \cite{Turrin1977,Wunderlich2007}. Finally, direct frequency-comb Raman spectroscopy of the two $|D_{5/2}, m_j=\pm 1/2\rangle \leftrightarrow |D_{3/2}, m_j'=\pm 1/2\rangle$ symmetric transitions is carried out \cite{Solaro2018}. The state of the ion is read out by the electron-shelving technique \cite{Dehmelt1982}. The first-order differential Zeeman shift induced by the static magnetic field is canceled by averaging the two transition frequencies. The differential AC-Stark shift induced by the frequency comb is reduced by taking advantage of the existence of a ``magic polarization'' \cite{Solaro2018}, and the unshifted transition frequency is obtained by extrapolating the measured frequencies to zero light intensity. The measured absolute D-splitting isotope shifts $\delta\nu^{A,40}_\text{DSIS}$ corrected for systematic effects (i.e. second-order Zeeman shift and electric-quadrupole shift mainly \cite{Solaro2018}) are presented table \ref{table:shifts}. The achieved relative accuracy ranges from 2 to \mbox{7 $\times 10^{-6}$}.

The 4s$~^2$S$_{1/2} \leftrightarrow 3$d$~^2$D$_{5/2}$ transition near 729 nm was measured by Rabi spectroscopy \cite{Rabi_&al_1939}. In an experimental sequence similar to the one described above, after the optical pumping stage, the two $|S_{1/2}, \pm1/2\rangle \leftrightarrow |D_{5/2}, \mp 3/2\rangle $ transitions are probed consecutively with $\pi$ pulses. The interrogation laser is locked to an ultra-stable high-finesse cavity providing a sub-kHz linewidth at short term (see figure \ref{fig:setup}). The absolute laser frequency is measured by beating this laser with one tooth of the frequency comb with the latter locked to an acetylene-stabilized ultra-stable fiber laser (\textit{Stabi$\lambda$aser} from Denmark's National Metrology Institute \cite{ stabilaser, Talvard2017}). The differential first-order Zeeman shift is once again canceled by averaging the two transition frequencies. These measurements are limited by the relative inaccuracy of our GPS-disciplined rubidium standard which was measured against the \textit{Stabi$\lambda$aser} to be $5\times 10^{-12}$. This corresponds to a 2 kHz accuracy on the S$_{1/2}$-D$_{5/2}$ transition and a relative accuracy on $\delta\nu^{A,40}_{729}$ ranging from 2 to \mbox{7 $\times 10^{-7}$}. The deduced isotope shifts $\delta\nu^{A,40}_{729}$ are given table \ref{table:shifts} together with part-per-billion measurements of $\delta\nu^{42,44,48-40}_{729}$ reported by Knollmann \textit{et al.} \cite{Knollmann2019}. Combined with our DSIS measurements, the data of Ref. \cite{Knollmann2019} are further used to calculate the isotope shifts of the 4s$~^2$S$_{1/2} \leftrightarrow 3$d$~^2$D$_{3/2}$ transition near 732 nm with a fractional accuracy better than $10^{-8}$, as also presented in table \ref{table:shifts}.

\begin{table}
\caption{\label{table:shifts} Isotope shifts relative to $^{40}$Ca$^+$ in MHz and their 1 standard deviation $\sigma$ uncertainties.}
\begin{ruledtabular}
\begin{tabular}{lllllrrr}
 A &\multicolumn{1}{c}{$\delta\nu^{A,40}_\text{DSIS}$} &\multicolumn{1}{c}{$\delta\nu^{A,40}_\text{729}$} &\multicolumn{1}{c}{$\delta\nu^{A,40}_\text{732}$\footnotemark[1]}\\
\hline
42& -3.519 896(24) & 2 771.873(2) & 2 775.393(2)  \\
& & 2 771.872 467 6(76)\footnotemark[2]  & 2 775.392 363(25)\footnotemark[3] \\ 
44& -6.792 470(22) & 5 340.888(2) & 5 347.680(2) \\
& & 5 340.887 394 6(78)\footnotemark[2] & 5 347.679 865(23)\footnotemark[3] \\
46& -9.901 524(21) & 7 768.401(2) & 7 778.302(2)  \\
& & ~~~~~~~/ & ~~~~~~~/ \\
48& -12.746 610(27) & 9 990.383(2) & 10 003.130(2)  \\
& & 9 990.381 870 0(63)\footnotemark[2] & 10 003.128 480(28)\footnotemark[3]   \\
\end{tabular}
\end{ruledtabular}
\footnotetext[1]{Calculated: $\delta\nu^{A,40}_\text{732} = \delta\nu^{A,40}_\text{729}-\delta\nu^{A,40}_\text{DSIS}$}
\footnotetext[2]{Taken from Ref. \cite{Knollmann2019}}
\footnotetext[3]{Calculated using values of $\delta\nu^{A,40}_\text{729}$ from Ref. \cite{Knollmann2019}.}
\end{table}

The two leading contributions to the isotope shift in atomic transition frequencies are the mass shift (MS) and the field shift (FS) \cite{2007a}. The MS originates from the difference of the nuclear mass which leads to differences in the nuclear recoil energy. The FS originates from the change in the effective nuclear charge radius, which leads to different electronic potentials near the origin. With these two contributions, the isotope shift of a transition $i$ between isotope $A$ and $A'$ can be written to leading order as:
\begin{equation}\label{eq:shifts}
\delta\nu_i^{AA'} \equiv \nu_i^{A}-\nu_i^{A'} = \delta\nu_{i,\text{MS}}^{AA'} + \delta\nu_{i,\text{FS}}^{AA'}
=  \frac{K_i}{\mu} + F_i \, \crad
\end{equation}
where $K_i$ and $F_i$ are the mass and field shift constants respectively, $\delta\left<r_c^2\right>^{AA'}=\left<r_c^2\right>^{A}-\left<r_c^2\right>^{A'}$ is the difference of the mean squared nuclear charge radii, and $\mu$ is the reduced mass given by \cite{Kurth1995}
\begin{equation}
\label{eq:mu}
\mu = \mu^{AA'} = \frac{m_{A'}(m_{A} + m_e)}{m_{A} - m_{A'}}
\end{equation}
where $m_e$ is the electron mass, and $m_A$ and $m_{A'}$ are the masses of the \emph{nuclei} of the two isotopes respectively. The nuclear masses can be deduced from the precisely determined masses of the neutral atomic calcium  isotopes~\cite{Wang2017}, the total mass of the electrons, and the sum of the electrons binding energies $E^\text{b}_n$: 
\begin{equation}
\label{eq:nuclear_mass}
	m_A = m_{A,\text{neutral atom}} - 20 m_e + \sum_{n=1}^{20} E^\text{b}_n %m_\text{binding energy, n}
\end{equation}
where the electron binding energies have been extracted from the NIST database \cite{kramida2018}. If the isotope shifts are measured for more than one transition, the equation \ref{eq:shifts} allows one to eliminate the typically poorly known \crad and to write the so-called King relation \cite{King1963}:
\begin{equation}\label{eq:king}
\begin{array}{ll}
\mu \delta\nu_{i}^{A A'} &= K_i - \frac{F_i}{F_j} K_j + \frac{F_i}{F_j} \, \mu \, \delta\nu_{j}^{A A'}\\
	%&= c + m\cdot \left( \mu \, \delta\nu_{j}^{A A'}\right)
\end{array}
\end{equation}
which, to leading order within the SM, is a linear relation between the modified isotope shifts $\mu\delta\nu_{i}^{A A'}$ and $\mu\delta\nu_{j}^{A A'}$ of the two transitions $i,\,j$. A NP interaction mediated by a boson $\phi$ of spin $s$ with coupling strengths $y_e$ and $y_n$ to electrons and neutrons, respectively, modifies the isotope shift predictions of Eq. \ref{eq:shifts} as
\begin{equation}\label{eq:NP}
\delta\nu_i^{AA'} =  \frac{K_i}{\mu} + F_i \, \crad \ + (-1)^s\frac{\hbar c}{4\pi}\frac{y_ey_n}{\hbar c}X_i\gamma^{AA'}
\end{equation}
where the electronic NP coefficient $X_i$ characterizes the overlap of the wave-functions of the lower and upper states of transition $i$ with the potential mediated by the boson, independent of the isotopes, and $\gamma^{AA'}$ depends on the isotopes only, independent of the transition. If $\phi$ couples linearly to the nucleus, then $\gamma^{AA'} = A-A'$. As a consequence, the King relation in Eq. \ref{eq:king} is in this case not linear anymore. Therefore, searching for non-linearities of the corresponding King plot provides sensitivity to a NP interaction mediated by such a boson.

\begin{figure}[t!]
	\centering
	\includegraphics[width=.48\textwidth]{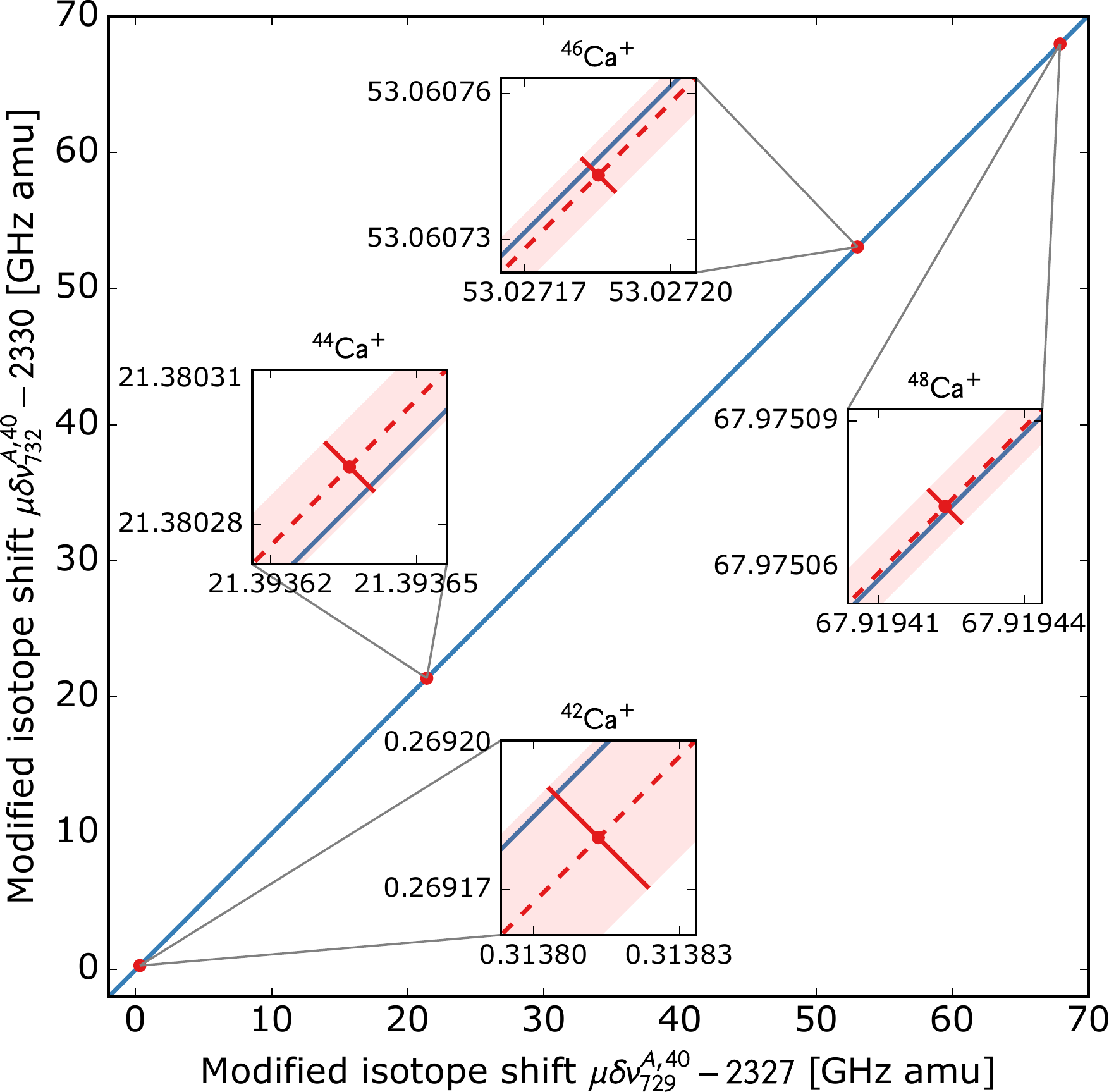}
	\caption{Two-dimensional King plot of the 732-nm and 729-nm transitions. The line is a fit to our data using a weighted orthogonal distance regression. The extracted fit parameters are given in the text. We point out that the isotope shift of the 732-nm transition is deduced from measurements of the isotope shift of the 729-nm transition $\delta\nu^{A,40}_\text{729}$ and of the D-splitting isotope shift $\delta\nu^{A,40}_\text{DSIS}$. Hence, the measurement accuracy on $\delta\nu^{A,40}_\text{729}$ ($\delta\nu^{A,40}_\text{DSIS}$) translates into an error bar parallel (perpendicular) to the fitted line, emphasizing that the analysis is limited by the achieved fractional accuracy on $\delta\nu^{A,40}_\text{DSIS}$.}
\label{fig:king_Dfine_729}
\end{figure}

The King plot of the modified isotope shift of the 732-nm transition against the modified isotope shift of the 729-nm transition, using our experimental data only, is shown figure \ref{fig:king_Dfine_729}. The blue line is a linear fit of the data using the King relation Eq. \ref{eq:king} and a Weighted Orthogonal Distance Regression \cite{Boggs1987}. We emphasize that $\delta\nu_{732}^{A,40}$ is deduced from measurements of $\delta\nu_{729}^{A,40}$ and $\delta\nu_\text{DSIS}^{A,40}$, and that $\delta\nu_{729}^{A,40} \gg \delta\nu_\text{DSIS}^{A,40}$. Consequently, the measurement uncertainties on $\delta\nu_{729}^{A,40}$ and $\delta\nu_\text{DSIS}^{A,40}$ translate into error bars essentially parallel and perpendicular to the fitted line, illustrating that the analysis is limited nearly exclusively by the achieved accuracy on $\delta\nu_\text{DSIS}^{A,40}$. 
In fact, as long as the fractional accuracy on $\delta\nu_{729}^{A,40}$ is smaller than the fractional accuracy on $\delta\nu_\text{DSIS}^{A,40}$, measuring the DSIS at, e.g., the 20 Hz level is equivalent to measuring both the 729-IS and the 732-IS with the same 20 Hz accuracy. This is a consequence of the King plot analysis being sensitive to the difference of isotope shifts of the D$_{3/2}$ and D$_{5/2}$ states, and this demonstrates the potential of measuring the DSIS directly using direct frequency-comb Raman spectroscopy.

\begin{figure}
	\centering
	\includegraphics[width=.48\textwidth]{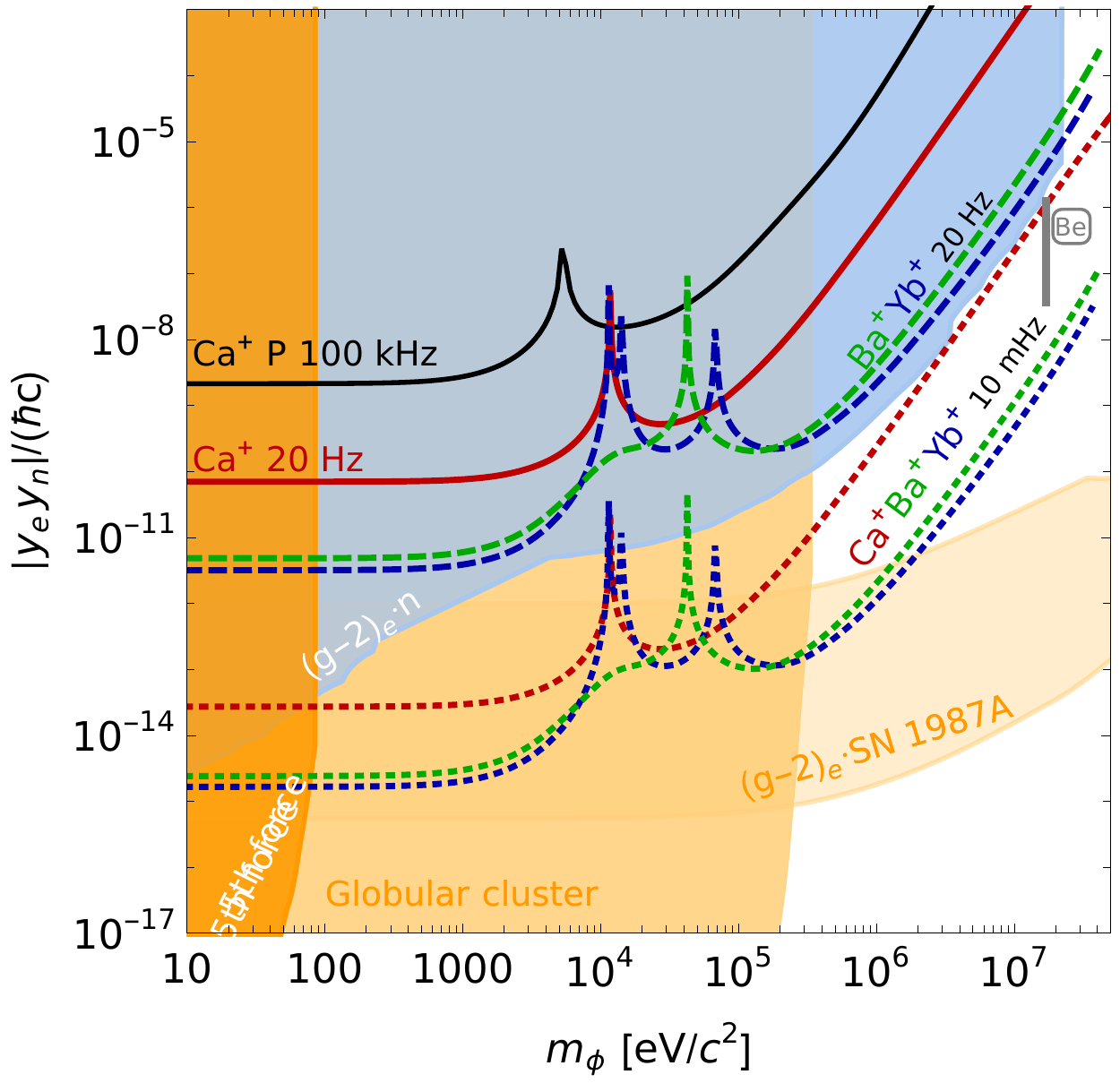}
	\caption{
	Current and projected constraints (2$\sigma$) on the coupling strength $y_ey_n$ of electrons and neutrons to a new boson $\phi$ of mass $m_\phi$. %(in natural units of $\hbar=c=1$\footnote{To recover dimensions: $y_ey_n \to y_ey_n/(\hbar c)$, $m_\phi/\ev \to m_\phi/(\ev/c^2)$}).  
    Existing bound \cite{Berengut2018} from measurements of the S$_{1/2}$-P$_{1/2}$ and D$_{3/2}$-P$_{1/2}$ transition isotope shifts in Ca$^+$ with an accuracy of $\mathcal{O}(100\,\text{kHz})$ \cite{Gebert2015} (black, labelled as P).
	Constraint imposed by this work (red, solid), limited by the $\sim$ 20\,Hz measurement uncertainty of the DSIS. Projection for a 10 mHz uncertainty on the DSIS (red, dotted).
	Projected constraints from measurements in Ba$^+$ (DSIS at 20\,Hz level, green dashed; 10\,mHz, green dotted) and in Yb$^+$ isotopes (dark blue, also for 20\,Hz and 10\,mHz) (for details see~\cite{supmat}). The curves end at $m_\phi$ corresonding to the corresponding inverse nuclear radii.
	For comparison, constraints from other experiments are shown as shaded areas \cite{Berengut2018}: fifth force~\cite{Bordag:2001qi,bordag2009advances} (dark orange), $(g-2)_e$ measurements~\cite{Olive:2016xmw,Hanneke:2010au} combined with neutron scattering data~\cite{Barbieri:1975xy,Leeb:1992qf,Pokotilovski:2006up,Nesvizhevsky:2007by} (light blue), or SN 1987A (light orange), and star cooling in globular clusters~\cite{Yao:2006px,Grifols:1986fc,Grifols:1988fv,Redondo:2013lna,Hardy:2016kme} (orange). The gray bar represents the range of $y_ey_n$ needed to explain the Be anomaly \cite{Krasznahorkay:2015iga,Feng:2016jff,Feng:2016ysn,Berengut2018,Frugiuele2017}.
		}
\label{fig:bounds}
\end{figure}

The reduced $\chi^2$ of the fit is 0.89 and the King plot is thus linear within our measurement uncertainty. The non-linearity (defined in the Supplemental Material \cite{[{See Supplemental Material at [URL will be inserted by publisher] for details regarding the calculation of the bounds}] supmat}) is $1.26\sigma$.
This allows for translating our measurement uncertainty into a constraint on the coupling strength of a hypothetical boson $\phi$. 
For a Yukawa potential %(in natural units, i.e., $\hbar=c=1$)
%\footnote{We work in natural units, $\hbar=c=1$, such that the potential 
%$V_{\text{NP}} = (-1)^s (A-Z)\hbar c\frac{y_e y_n}{4\pi \hbar c} \frac{e^{-m_\phi r c/\hbar}}{r}$ is reduced to the one in the text.},
$V_{\text{NP}} = (-1)^s (A-Z)\frac{\hbar c}{4\pi}\frac{y_e y_n}{\hbar c} \frac{e^{- r m_\phi c/\hbar}}{r}$,
where $Z$ is the number of protons,
we calculate the electronic NP coefficients $X_i$ using Brueckner orbitals and including relativistic random phase approximation corrections to the operator (see \cite{supmat}). By constraining the nonlinear term from data (see \cite{supmat,Berengut2018}) and using the theory calculation of $X_i$, we evaluate the bounds on $y_ey_n$ as a function of the new mediator's mass $m_\phi$ which are shown in Fig.~\ref{fig:bounds}. 
The red solid curve corresponds to the bound using our experimental data only, yielding $y_ey_n/\hbar c<6.9\cdot 10^{-11}$ at the $2\sigma$ level in the mass-less limit ($m_\phi=1\ev$).
The combination of the 729-IS measurements of Ref. \cite{Knollmann2019} with our measurements of the DSIS and of $\delta\nu_{729}^{40,46}$, however, does not improve the bound despite the thousand times better accuracy on $\delta\nu_{729}^{42,44,48-40}$, confirming that the accuracy on the DSIS is the limiting one 
(as long as $\sigma_{S-D_{5/2}}\cdot F_{\rm DSIS}/F_{S-D_{5/2}} < \sigma_{\rm DSIS}$) 
and illustrating the potential of measuring the DSIS directly. 
The combined bound coincides with the bound using purely our data (up to a relative difference of 1\%) and is therefore not displayed.
The black curve corresponds to the previous best bound set by measurements of the isotope shift of the two S$_{1/2}$-P$_{1/2}$ and D$_{3/2}$-P$_{1/2}$ dipole-allowed transitions by Gebert \textit{et al.} \cite{Gebert2015} limiting 
$y_ey_n/\hbar c<2\cdot 10^{-9}$ for $m_\phi=0$.
We note that despite the hundred times better relative accuracy on the two 4s - 3d transition isotope shifts achieved in this work, the bounds on $y_ey_n$ are improved by less than a factor 100. This is because the electronic configurations of the D$_{3/2}$ and D$_{5/2}$ states are more similar than the ones of the relevant S$_{1/2}$ and D$_{3/2}$ states of Ref. \cite{Gebert2015}. 
More stringent bounds could be placed by constraining King plot non-linearities with heavier elements provided that one can correct for the non-linearities already predicted at higher order within the SM \cite{Flambaum2018}. 
Two promising elements are Ba$^+$ or Yb$^+$ which both have five spin-0 isotopes and D-splittings of 24 and 42 THz, respectively. The projected constraints imposed by measuring the DSIS at the 20 Hz level and the S$_{1/2}$-D$_{5/2}$ transition isotope shifts at the kHz level in Ba$^+$ (green, dashed) and Yb$^+$ (dark blue, dashed) are also plotted in Fig. \ref{fig:bounds} (see \cite{supmat}). Furthermore, we estimate the sensitivity of Ca$^+$, Ba$^+$ and Yb$^+$ for measurements of the DSIS with 10 mHz accuracy and of the S$_{1/2}$-D$_{5/2}$ transition isotope shifts with $\sim$ Hz accuracy, under the condition that the uncertainty is limited by the isotope shift measurements and not by the uncertainty on the masses. 
The current constraints on $y_ey_n$ from King plot analyses, included the new bound derived in this work, are weaker than the astrophysical bound from star cooling of globular clusters~\cite{Yao:2006px,Grifols:1986fc,Grifols:1988fv,Redondo:2013lna,Hardy:2016kme} for $m_\phi\lesssim 0.3$\,MeV/$c^2$ and weaker than constraint on $y_e$ from the magnetic dipole moment $(g-2)$ of the electron~\cite{Olive:2016xmw,Hanneke:2010au} combined with the constraint on $y_n$ from neutron scattering~\cite{Barbieri:1975xy,Leeb:1992qf,Pokotilovski:2006up,Nesvizhevsky:2007by}. In contrast, the improved accuracy of the DSIS and S$_{1/2}$-D$_{5/2}$ measurements have the potential to probe so far unconstrained parameter space for $m_\phi\gtrsim 0.3$\,MeV/$c^2$ and in particular the range of $y_ey_n$ at $m_\phi=17$ MeV/$c^2$ needed to explain the Be anomaly.

Finally, considering the case without a NP contribution, 
the fit parameters of the King plot analysis are $K_{21}=K_\text{732} - F_\text{732}/F_\text{729} K_\text{729} = -0.4961(5)$ GHz.amu and $F_{21}=F_\text{732}/F_\text{729} = 1.00148305(20)$. Notably, we extract the ratio of the field shift constants with a relative accuracy of $2\times 10^{-7}$ and the obtained value matches well the theoretical value calculated using many-body perturbation theory (see \cite{supmat}) $F^\mathrm{MBPT}_{21}=1.0016$. We mention that this is also the case for the field shift ratio of the S$_{1/2}$-P$_{1/2}$ and S$_{1/2}$-P$_{3/2}$ transitions which isotope shifts were recently measured by Müller \textit{et al.} \cite{Muller2020}, solving the field shift puzzle introduced with previous measurements made by Shi \textit{et al.} \cite{Shi2016}. Lastly, our data could be used to improve the accuracy on \crad for the even calcium isotopes considered here \cite{Kramida2020}.

In summary, we have reported measurements of the D-fine-structure splitting isotope shift using direct frequency-comb Raman spectroscopy on all stable even isotopes of $^A$Ca$^+$ (A = 40,\,42,\,44,\,46, and 48) with an accuracy of $\sim 20$ Hz. Combined with isotope shift measurements of the 4s$~^2$S$_{1/2}  \leftrightarrow 3$d$~^2$D$_{5/2}$ transition at the 2 kHz level, we performed a King plot analysis of the 4s$~^2$S$_{1/2}  \leftrightarrow 3$d$~^2$D$_{5/2}$ and 4s$~^2$S$_{1/2}  \leftrightarrow 3$d$~^2$D$_{3/2}$ transitions with unprecedented accuracy and extracted the field shift ratio with a fractional accuracy of $2\times 10^{-7}$. Furthermore, the achieved uncertainty on the King plot linearity was used to improve isotope-shift-based bounds on the coupling strength of a New Physics boson to both electrons and neutrons. More stringent bounds could be placed by looking for King plot non-linearities with heavier elements, such as Ba$^+$ or Yb$^+$ applying direct frequency-comb Raman spectroscopy. Finally, NP interactions mediated by bosons with masses $m_\phi\gtrsim 0.3$\,MeV/$c^2$ and so far unconstrained by experiments, could be probed by measuring, with existing techniques for optical atomic clocks, the DSIS at the 10 mHz level and one of the S-D isotope shifts at the $\sim$ Hz level.

\begin{acknowledgments}
We thank J. Hur, V. Vuleti\'c, M. Schlaffer, R. Ozeri and P. O. Schmidt for interesting discussions. We acknowledge the Danish Center for Laser Infrastructure, LASERLAB.DK, established through the support of the Danish Ministry of Research and Education, for financial support and access to the frequency comb. We also acknowledge support from Innovation Fund Denmark, through the Quantum Innovation Center, Qubiz, for financial support and access to the \textit{Stabi$\lambda$aser}. C. S. acknowledges support from the European Commission through the Marie Curie Individual Fellowship COMAMOC (grant agreement no 795107) under Horizon 2020. S. M. and K. F. acknowledge support from the European Commission through the Marie Curie Initial Training Network COMIQ (grant agreement no 607491) under FP7. 
E. F. was supported during part of this work by the Minerva Foundation.
J. C. B. is supported in this work by the Australian Research Council (DP190100974). M. D. acknowledges support from the European Commission’s FET Open TEQ, the Villum Foundation and the Sapere Aude Initiative from the Independent Research Fund Denmark.
\end{acknowledgments}

\bibliography{IsotopesNew}

\begin{thebibliography}{53}%
\makeatletter
\providecommand \@ifxundefined [1]{%
 \@ifx{#1\undefined}
}%
\providecommand \@ifnum [1]{%
 \ifnum #1\expandafter \@firstoftwo
 \else \expandafter \@secondoftwo
 \fi
}%
\providecommand \@ifx [1]{%
 \ifx #1\expandafter \@firstoftwo
 \else \expandafter \@secondoftwo
 \fi
}%
\providecommand \natexlab [1]{#1}%
\providecommand \enquote  [1]{``#1''}%
\providecommand \bibnamefont  [1]{#1}%
\providecommand \bibfnamefont [1]{#1}%
\providecommand \citenamefont [1]{#1}%
\providecommand \href@noop [0]{\@secondoftwo}%
\providecommand \href [0]{\begingroup \@sanitize@url \@href}%
\providecommand \@href[1]{\@@startlink{#1}\@@href}%
\providecommand \@@href[1]{\endgroup#1\@@endlink}%
\providecommand \@sanitize@url [0]{\catcode `\\12\catcode `\$12\catcode
  `\&12\catcode `\#12\catcode `\^12\catcode `\_12\catcode `\%12\relax}%
\providecommand \@@startlink[1]{}%
\providecommand \@@endlink[0]{}%
\providecommand \url  [0]{\begingroup\@sanitize@url \@url }%
\providecommand \@url [1]{\endgroup\@href {#1}{\urlprefix }}%
\providecommand \urlprefix  [0]{URL }%
\providecommand \Eprint [0]{\href }%
\providecommand \doibase [0]{http://dx.doi.org/}%
\providecommand \selectlanguage [0]{\@gobble}%
\providecommand \bibinfo  [0]{\@secondoftwo}%
\providecommand \bibfield  [0]{\@secondoftwo}%
\providecommand \translation [1]{[#1]}%
\providecommand \BibitemOpen [0]{}%
\providecommand \bibitemStop [0]{}%
\providecommand \bibitemNoStop [0]{.\EOS\space}%
\providecommand \EOS [0]{\spacefactor3000\relax}%
\providecommand \BibitemShut  [1]{\csname bibitem#1\endcsname}%
\let\auto@bib@innerbib\@empty
%</preamble>
\bibitem [{\citenamefont {Knollmann}\ \emph {et~al.}(2019)\citenamefont
  {Knollmann}, \citenamefont {Patel},\ and\ \citenamefont
  {Doret}}]{Knollmann2019}%
  \BibitemOpen
  \bibfield  {author} {\bibinfo {author} {\bibfnamefont {F.~W.}\ \bibnamefont
  {Knollmann}}, \bibinfo {author} {\bibfnamefont {A.~N.}\ \bibnamefont
  {Patel}}, \ and\ \bibinfo {author} {\bibfnamefont {S.~C.}\ \bibnamefont
  {Doret}},\ }\href {\doibase 10.1103/physreva.100.022514} {\bibfield
  {journal} {\bibinfo  {journal} {Physical Review A}\ }\textbf {\bibinfo
  {volume} {100}} (\bibinfo {year} {2019}),\
  10.1103/physreva.100.022514}\BibitemShut {NoStop}%
\bibitem [{\citenamefont {Mangano}(2020)}]{Mangano:2020icy}%
  \BibitemOpen
  \bibfield  {author} {\bibinfo {author} {\bibfnamefont {M.}~\bibnamefont
  {Mangano}},\ }\href@noop {} {\bibfield  {journal} {\bibinfo  {journal} {arXiv
  preprint arXiv:2003.05976}\ } (\bibinfo {year} {2020})}\BibitemShut {NoStop}%
\bibitem [{\citenamefont {BETTONI}\ \emph {et~al.}(2006)\citenamefont
  {BETTONI}, \citenamefont {BIANCO}, \citenamefont {BOSSI}, \citenamefont
  {CATANESI}, \citenamefont {CECCUCCI}, \citenamefont {CERVELLI}, \citenamefont
  {DELLORSO}, \citenamefont {DOSSELLI}, \citenamefont {FERRONI},\ and\
  \citenamefont {GRASSI}}]{Bettoni:2006za}%
  \BibitemOpen
  \bibfield  {author} {\bibinfo {author} {\bibfnamefont {D.}~\bibnamefont
  {BETTONI}}, \bibinfo {author} {\bibfnamefont {S.}~\bibnamefont {BIANCO}},
  \bibinfo {author} {\bibfnamefont {F.}~\bibnamefont {BOSSI}}, \bibinfo
  {author} {\bibfnamefont {M.}~\bibnamefont {CATANESI}}, \bibinfo {author}
  {\bibfnamefont {A.}~\bibnamefont {CECCUCCI}}, \bibinfo {author}
  {\bibfnamefont {F.}~\bibnamefont {CERVELLI}}, \bibinfo {author}
  {\bibfnamefont {M.}~\bibnamefont {DELLORSO}}, \bibinfo {author}
  {\bibfnamefont {U.}~\bibnamefont {DOSSELLI}}, \bibinfo {author}
  {\bibfnamefont {F.}~\bibnamefont {FERRONI}}, \ and\ \bibinfo {author}
  {\bibfnamefont {M.}~\bibnamefont {GRASSI}},\ }\href {\doibase
  10.1016/j.physrep.2006.07.003} {\bibfield  {journal} {\bibinfo  {journal}
  {Physics Reports}\ }\textbf {\bibinfo {volume} {434}},\ \bibinfo {pages} {47}
  (\bibinfo {year} {2006})}\BibitemShut {NoStop}%
\bibitem [{\citenamefont {DeMille}\ \emph {et~al.}(2017)\citenamefont
  {DeMille}, \citenamefont {Doyle},\ and\ \citenamefont
  {Sushkov}}]{DeMille:2017azs}%
  \BibitemOpen
  \bibfield  {author} {\bibinfo {author} {\bibfnamefont {D.}~\bibnamefont
  {DeMille}}, \bibinfo {author} {\bibfnamefont {J.~M.}\ \bibnamefont {Doyle}},
  \ and\ \bibinfo {author} {\bibfnamefont {A.~O.}\ \bibnamefont {Sushkov}},\
  }\href {\doibase 10.1126/science.aal3003} {\bibfield  {journal} {\bibinfo
  {journal} {Science}\ }\textbf {\bibinfo {volume} {357}},\ \bibinfo {pages}
  {990} (\bibinfo {year} {2017})}\BibitemShut {NoStop}%
\bibitem [{\citenamefont {Safronova}\ \emph {et~al.}(2018)\citenamefont
  {Safronova}, \citenamefont {Budker}, \citenamefont {DeMille}, \citenamefont
  {Kimball}, \citenamefont {Derevianko},\ and\ \citenamefont
  {Clark}}]{Safronova2018}%
  \BibitemOpen
  \bibfield  {author} {\bibinfo {author} {\bibfnamefont {M.}~\bibnamefont
  {Safronova}}, \bibinfo {author} {\bibfnamefont {D.}~\bibnamefont {Budker}},
  \bibinfo {author} {\bibfnamefont {D.}~\bibnamefont {DeMille}}, \bibinfo
  {author} {\bibfnamefont {D.~F.~J.}\ \bibnamefont {Kimball}}, \bibinfo
  {author} {\bibfnamefont {A.}~\bibnamefont {Derevianko}}, \ and\ \bibinfo
  {author} {\bibfnamefont {C.~W.}\ \bibnamefont {Clark}},\ }\href {\doibase
  10.1103/revmodphys.90.025008} {\bibfield  {journal} {\bibinfo  {journal}
  {Reviews of Modern Physics}\ }\textbf {\bibinfo {volume} {90}} (\bibinfo
  {year} {2018}),\ 10.1103/revmodphys.90.025008}\BibitemShut {NoStop}%
\bibitem [{\citenamefont {Delaunay}\ \emph
  {et~al.}(2017{\natexlab{a}})\citenamefont {Delaunay}, \citenamefont
  {Frugiuele}, \citenamefont {Fuchs},\ and\ \citenamefont
  {Soreq}}]{Delaunay2017b}%
  \BibitemOpen
  \bibfield  {author} {\bibinfo {author} {\bibfnamefont {C.}~\bibnamefont
  {Delaunay}}, \bibinfo {author} {\bibfnamefont {C.}~\bibnamefont {Frugiuele}},
  \bibinfo {author} {\bibfnamefont {E.}~\bibnamefont {Fuchs}}, \ and\ \bibinfo
  {author} {\bibfnamefont {Y.}~\bibnamefont {Soreq}},\ }\href {\doibase
  10.1103/physrevd.96.115002} {\bibfield  {journal} {\bibinfo  {journal}
  {Physical Review D}\ }\textbf {\bibinfo {volume} {96}} (\bibinfo {year}
  {2017}{\natexlab{a}}),\ 10.1103/physrevd.96.115002}\BibitemShut {NoStop}%
\bibitem [{\citenamefont {Delaunay}\ \emph
  {et~al.}(2017{\natexlab{b}})\citenamefont {Delaunay}, \citenamefont {Ozeri},
  \citenamefont {Perez},\ and\ \citenamefont {Soreq}}]{Delaunay2017a}%
  \BibitemOpen
  \bibfield  {author} {\bibinfo {author} {\bibfnamefont {C.}~\bibnamefont
  {Delaunay}}, \bibinfo {author} {\bibfnamefont {R.}~\bibnamefont {Ozeri}},
  \bibinfo {author} {\bibfnamefont {G.}~\bibnamefont {Perez}}, \ and\ \bibinfo
  {author} {\bibfnamefont {Y.}~\bibnamefont {Soreq}},\ }\href {\doibase
  10.1103/physrevd.96.093001} {\bibfield  {journal} {\bibinfo  {journal}
  {Physical Review D}\ }\textbf {\bibinfo {volume} {96}} (\bibinfo {year}
  {2017}{\natexlab{b}}),\ 10.1103/physrevd.96.093001}\BibitemShut {NoStop}%
\bibitem [{\citenamefont {King}(1963)}]{King1963}%
  \BibitemOpen
  \bibfield  {author} {\bibinfo {author} {\bibfnamefont {W.~H.}\ \bibnamefont
  {King}},\ }\href {\doibase 10.1364/josa.53.000638} {\bibfield  {journal}
  {\bibinfo  {journal} {Journal of the Optical Society of America}\ }\textbf
  {\bibinfo {volume} {53}},\ \bibinfo {pages} {638} (\bibinfo {year}
  {1963})}\BibitemShut {NoStop}%
\bibitem [{\citenamefont {Berengut}\ \emph {et~al.}(2018)\citenamefont
  {Berengut}, \citenamefont {Budker}, \citenamefont {Delaunay}, \citenamefont
  {Flambaum}, \citenamefont {Frugiuele}, \citenamefont {Fuchs}, \citenamefont
  {Grojean}, \citenamefont {Harnik}, \citenamefont {Ozeri}, \citenamefont
  {Perez},\ and\ \citenamefont {Soreq}}]{Berengut2018}%
  \BibitemOpen
  \bibfield  {author} {\bibinfo {author} {\bibfnamefont {J.~C.}\ \bibnamefont
  {Berengut}}, \bibinfo {author} {\bibfnamefont {D.}~\bibnamefont {Budker}},
  \bibinfo {author} {\bibfnamefont {C.}~\bibnamefont {Delaunay}}, \bibinfo
  {author} {\bibfnamefont {V.~V.}\ \bibnamefont {Flambaum}}, \bibinfo {author}
  {\bibfnamefont {C.}~\bibnamefont {Frugiuele}}, \bibinfo {author}
  {\bibfnamefont {E.}~\bibnamefont {Fuchs}}, \bibinfo {author} {\bibfnamefont
  {C.}~\bibnamefont {Grojean}}, \bibinfo {author} {\bibfnamefont
  {R.}~\bibnamefont {Harnik}}, \bibinfo {author} {\bibfnamefont
  {R.}~\bibnamefont {Ozeri}}, \bibinfo {author} {\bibfnamefont
  {G.}~\bibnamefont {Perez}}, \ and\ \bibinfo {author} {\bibfnamefont
  {Y.}~\bibnamefont {Soreq}},\ }\href {\doibase 10.1103/physrevlett.120.091801}
  {\bibfield  {journal} {\bibinfo  {journal} {Physical Review Letters}\
  }\textbf {\bibinfo {volume} {120}} (\bibinfo {year} {2018}),\
  10.1103/physrevlett.120.091801}\BibitemShut {NoStop}%
\bibitem [{\citenamefont {Frugiuele}\ \emph {et~al.}(2017)\citenamefont
  {Frugiuele}, \citenamefont {Fuchs}, \citenamefont {Perez},\ and\
  \citenamefont {Schlaffer}}]{Frugiuele2017}%
  \BibitemOpen
  \bibfield  {author} {\bibinfo {author} {\bibfnamefont {C.}~\bibnamefont
  {Frugiuele}}, \bibinfo {author} {\bibfnamefont {E.}~\bibnamefont {Fuchs}},
  \bibinfo {author} {\bibfnamefont {G.}~\bibnamefont {Perez}}, \ and\ \bibinfo
  {author} {\bibfnamefont {M.}~\bibnamefont {Schlaffer}},\ }\href {\doibase
  10.1103/physrevd.96.015011} {\bibfield  {journal} {\bibinfo  {journal}
  {Physical Review D}\ }\textbf {\bibinfo {volume} {96}} (\bibinfo {year}
  {2017}),\ 10.1103/physrevd.96.015011}\BibitemShut {NoStop}%
\bibitem [{\citenamefont {Feng}\ \emph {et~al.}(2016)\citenamefont {Feng},
  \citenamefont {Fornal}, \citenamefont {Galon}, \citenamefont {Gardner},
  \citenamefont {Smolinsky}, \citenamefont {Tait},\ and\ \citenamefont
  {Tanedo}}]{Feng:2016jff}%
  \BibitemOpen
  \bibfield  {author} {\bibinfo {author} {\bibfnamefont {J.~L.}\ \bibnamefont
  {Feng}}, \bibinfo {author} {\bibfnamefont {B.}~\bibnamefont {Fornal}},
  \bibinfo {author} {\bibfnamefont {I.}~\bibnamefont {Galon}}, \bibinfo
  {author} {\bibfnamefont {S.}~\bibnamefont {Gardner}}, \bibinfo {author}
  {\bibfnamefont {J.}~\bibnamefont {Smolinsky}}, \bibinfo {author}
  {\bibfnamefont {T.~M.~P.}\ \bibnamefont {Tait}}, \ and\ \bibinfo {author}
  {\bibfnamefont {P.}~\bibnamefont {Tanedo}},\ }\href {\doibase
  10.1103/PhysRevLett.117.071803} {\bibfield  {journal} {\bibinfo  {journal}
  {Phys.\ Rev.\ Lett.}\ }\textbf {\bibinfo {volume} {117}},\ \bibinfo {pages}
  {071803} (\bibinfo {year} {2016})},\ \Eprint
  {http://arxiv.org/abs/1604.07411} {arXiv:1604.07411 [hep-ph]} \BibitemShut
  {NoStop}%
\bibitem [{\citenamefont {Feng}\ \emph {et~al.}(2017)\citenamefont {Feng},
  \citenamefont {Fornal}, \citenamefont {Galon}, \citenamefont {Gardner},
  \citenamefont {Smolinsky}, \citenamefont {Tait},\ and\ \citenamefont
  {Tanedo}}]{Feng:2016ysn}%
  \BibitemOpen
  \bibfield  {author} {\bibinfo {author} {\bibfnamefont {J.~L.}\ \bibnamefont
  {Feng}}, \bibinfo {author} {\bibfnamefont {B.}~\bibnamefont {Fornal}},
  \bibinfo {author} {\bibfnamefont {I.}~\bibnamefont {Galon}}, \bibinfo
  {author} {\bibfnamefont {S.}~\bibnamefont {Gardner}}, \bibinfo {author}
  {\bibfnamefont {J.}~\bibnamefont {Smolinsky}}, \bibinfo {author}
  {\bibfnamefont {T.~M.~P.}\ \bibnamefont {Tait}}, \ and\ \bibinfo {author}
  {\bibfnamefont {P.}~\bibnamefont {Tanedo}},\ }\href {\doibase
  10.1103/PhysRevD.95.035017} {\bibfield  {journal} {\bibinfo  {journal}
  {Phys.\ Rev.\ D}\ }\textbf {\bibinfo {volume} {95}},\ \bibinfo {pages}
  {035017} (\bibinfo {year} {2017})},\ \Eprint
  {http://arxiv.org/abs/1608.03591} {arXiv:1608.03591 [hep-ph]} \BibitemShut
  {NoStop}%
\bibitem [{\citenamefont {Krasznahorkay}\ \emph {et~al.}(2016)\citenamefont
  {Krasznahorkay}, \citenamefont {Csatl{\'{o}}s}, \citenamefont {Csige},
  \citenamefont {G{\'{a}}csi}, \citenamefont {Guly{\'{a}}s}, \citenamefont
  {Hunyadi}, \citenamefont {Kuti}, \citenamefont {Nyak{\'{o}}}, \citenamefont
  {Stuhl}, \citenamefont {Tim{\'{a}}r}, \citenamefont {Tornyi}, \citenamefont
  {Vajta}, \citenamefont {Ketel},\ and\ \citenamefont
  {Krasznahorkay}}]{Krasznahorkay:2015iga}%
  \BibitemOpen
  \bibfield  {author} {\bibinfo {author} {\bibfnamefont {A.}~\bibnamefont
  {Krasznahorkay}}, \bibinfo {author} {\bibfnamefont {M.}~\bibnamefont
  {Csatl{\'{o}}s}}, \bibinfo {author} {\bibfnamefont {L.}~\bibnamefont
  {Csige}}, \bibinfo {author} {\bibfnamefont {Z.}~\bibnamefont {G{\'{a}}csi}},
  \bibinfo {author} {\bibfnamefont {J.}~\bibnamefont {Guly{\'{a}}s}}, \bibinfo
  {author} {\bibfnamefont {M.}~\bibnamefont {Hunyadi}}, \bibinfo {author}
  {\bibfnamefont {I.}~\bibnamefont {Kuti}}, \bibinfo {author} {\bibfnamefont
  {B.}~\bibnamefont {Nyak{\'{o}}}}, \bibinfo {author} {\bibfnamefont
  {L.}~\bibnamefont {Stuhl}}, \bibinfo {author} {\bibfnamefont
  {J.}~\bibnamefont {Tim{\'{a}}r}}, \bibinfo {author} {\bibfnamefont
  {T.}~\bibnamefont {Tornyi}}, \bibinfo {author} {\bibfnamefont
  {Z.}~\bibnamefont {Vajta}}, \bibinfo {author} {\bibfnamefont
  {T.}~\bibnamefont {Ketel}}, \ and\ \bibinfo {author} {\bibfnamefont
  {A.}~\bibnamefont {Krasznahorkay}},\ }\href {\doibase
  10.1103/physrevlett.116.042501} {\bibfield  {journal} {\bibinfo  {journal}
  {Physical Review Letters}\ }\textbf {\bibinfo {volume} {116}} (\bibinfo
  {year} {2016}),\ 10.1103/physrevlett.116.042501}\BibitemShut {NoStop}%
\bibitem [{\citenamefont {Gebert}\ \emph {et~al.}(2015)\citenamefont {Gebert},
  \citenamefont {Wan}, \citenamefont {Wolf}, \citenamefont {Angstmann},
  \citenamefont {Berengut},\ and\ \citenamefont {Schmidt}}]{Gebert2015}%
  \BibitemOpen
  \bibfield  {author} {\bibinfo {author} {\bibfnamefont {F.}~\bibnamefont
  {Gebert}}, \bibinfo {author} {\bibfnamefont {Y.}~\bibnamefont {Wan}},
  \bibinfo {author} {\bibfnamefont {F.}~\bibnamefont {Wolf}}, \bibinfo {author}
  {\bibfnamefont {C.~N.}\ \bibnamefont {Angstmann}}, \bibinfo {author}
  {\bibfnamefont {J.~C.}\ \bibnamefont {Berengut}}, \ and\ \bibinfo {author}
  {\bibfnamefont {P.~O.}\ \bibnamefont {Schmidt}},\ }\href {\doibase
  10.1103/PhysRevLett.115.053003} {\bibfield  {journal} {\bibinfo  {journal}
  {Phys. Rev. Lett.}\ }\textbf {\bibinfo {volume} {115}},\ \bibinfo {pages}
  {053003} (\bibinfo {year} {2015})}\BibitemShut {NoStop}%
\bibitem [{\citenamefont {Manovitz}\ \emph {et~al.}(2019)\citenamefont
  {Manovitz}, \citenamefont {Shaniv}, \citenamefont {Shapira}, \citenamefont
  {Ozeri},\ and\ \citenamefont {Akerman}}]{Manovitz2019}%
  \BibitemOpen
  \bibfield  {author} {\bibinfo {author} {\bibfnamefont {T.}~\bibnamefont
  {Manovitz}}, \bibinfo {author} {\bibfnamefont {R.}~\bibnamefont {Shaniv}},
  \bibinfo {author} {\bibfnamefont {Y.}~\bibnamefont {Shapira}}, \bibinfo
  {author} {\bibfnamefont {R.}~\bibnamefont {Ozeri}}, \ and\ \bibinfo {author}
  {\bibfnamefont {N.}~\bibnamefont {Akerman}},\ }\href {\doibase
  10.1103/physrevlett.123.203001} {\bibfield  {journal} {\bibinfo  {journal}
  {Physical Review Letters}\ }\textbf {\bibinfo {volume} {123}} (\bibinfo
  {year} {2019}),\ 10.1103/physrevlett.123.203001}\BibitemShut {NoStop}%
\bibitem [{\citenamefont {Chwalla}\ \emph {et~al.}(2009)\citenamefont
  {Chwalla}, \citenamefont {Benhelm}, \citenamefont {Kim}, \citenamefont
  {Kirchmair}, \citenamefont {Monz}, \citenamefont {Riebe}, \citenamefont
  {Schindler}, \citenamefont {Villar}, \citenamefont {H{\"a}nsel},
  \citenamefont {Roos} \emph {et~al.}}]{Chwalla2009}%
  \BibitemOpen
  \bibfield  {author} {\bibinfo {author} {\bibfnamefont {M.}~\bibnamefont
  {Chwalla}}, \bibinfo {author} {\bibfnamefont {J.}~\bibnamefont {Benhelm}},
  \bibinfo {author} {\bibfnamefont {K.}~\bibnamefont {Kim}}, \bibinfo {author}
  {\bibfnamefont {G.}~\bibnamefont {Kirchmair}}, \bibinfo {author}
  {\bibfnamefont {T.}~\bibnamefont {Monz}}, \bibinfo {author} {\bibfnamefont
  {M.}~\bibnamefont {Riebe}}, \bibinfo {author} {\bibfnamefont
  {P.}~\bibnamefont {Schindler}}, \bibinfo {author} {\bibfnamefont
  {A.}~\bibnamefont {Villar}}, \bibinfo {author} {\bibfnamefont
  {W.}~\bibnamefont {H{\"a}nsel}}, \bibinfo {author} {\bibfnamefont
  {C.}~\bibnamefont {Roos}},  \emph {et~al.},\ }\href {\doibase
  10.1103/PhysRevLett.102.023002} {\bibfield  {journal} {\bibinfo  {journal}
  {Physical review letters}\ }\textbf {\bibinfo {volume} {102}},\ \bibinfo
  {pages} {023002} (\bibinfo {year} {2009})}\BibitemShut {NoStop}%
\bibitem [{\citenamefont {Dehmelt}(1982)}]{Dehmelt1982}%
  \BibitemOpen
  \bibfield  {author} {\bibinfo {author} {\bibfnamefont {H.~G.}\ \bibnamefont
  {Dehmelt}},\ }\href {\doibase 10.1109/tim.1982.6312526} {\bibfield  {journal}
  {\bibinfo  {journal} {{IEEE} Transactions on Instrumentation and
  Measurement}\ }\textbf {\bibinfo {volume} {{IM}-31}},\ \bibinfo {pages} {83}
  (\bibinfo {year} {1982})}\BibitemShut {NoStop}%
\bibitem [{\citenamefont {Solaro}\ \emph {et~al.}(2018)\citenamefont {Solaro},
  \citenamefont {Meyer}, \citenamefont {Fisher}, \citenamefont {DePalatis},\
  and\ \citenamefont {Drewsen}}]{Solaro2018}%
  \BibitemOpen
  \bibfield  {author} {\bibinfo {author} {\bibfnamefont {C.}~\bibnamefont
  {Solaro}}, \bibinfo {author} {\bibfnamefont {S.}~\bibnamefont {Meyer}},
  \bibinfo {author} {\bibfnamefont {K.}~\bibnamefont {Fisher}}, \bibinfo
  {author} {\bibfnamefont {M.}~\bibnamefont {DePalatis}}, \ and\ \bibinfo
  {author} {\bibfnamefont {M.}~\bibnamefont {Drewsen}},\ }\href {\doibase
  10.1103/physrevlett.120.253601} {\bibfield  {journal} {\bibinfo  {journal}
  {Physical Review Letters}\ }\textbf {\bibinfo {volume} {120}} (\bibinfo
  {year} {2018}),\ 10.1103/physrevlett.120.253601}\BibitemShut {NoStop}%
\bibitem [{\citenamefont {Kj{\ae}rgaard}\ \emph {et~al.}(2000)\citenamefont
  {Kj{\ae}rgaard}, \citenamefont {Hornek{\ae}r}, \citenamefont {Thommesen},
  \citenamefont {Videsen},\ and\ \citenamefont {Drewsen}}]{Kjergaard2000}%
  \BibitemOpen
  \bibfield  {author} {\bibinfo {author} {\bibfnamefont {N.}~\bibnamefont
  {Kj{\ae}rgaard}}, \bibinfo {author} {\bibfnamefont {L.}~\bibnamefont
  {Hornek{\ae}r}}, \bibinfo {author} {\bibfnamefont {A.}~\bibnamefont
  {Thommesen}}, \bibinfo {author} {\bibfnamefont {Z.}~\bibnamefont {Videsen}},
  \ and\ \bibinfo {author} {\bibfnamefont {M.}~\bibnamefont {Drewsen}},\ }\href
  {\doibase 10.1007/s003400000296} {\bibfield  {journal} {\bibinfo  {journal}
  {Applied Physics B: Lasers and Optics}\ }\textbf {\bibinfo {volume} {71}},\
  \bibinfo {pages} {207} (\bibinfo {year} {2000})}\BibitemShut {NoStop}%
\bibitem [{\citenamefont {Mortensen}\ \emph {et~al.}(2004)\citenamefont
  {Mortensen}, \citenamefont {Lindballe}, \citenamefont {Jensen}, \citenamefont
  {Staanum}, \citenamefont {Voigt},\ and\ \citenamefont
  {Drewsen}}]{Mortensen2004}%
  \BibitemOpen
  \bibfield  {author} {\bibinfo {author} {\bibfnamefont {A.}~\bibnamefont
  {Mortensen}}, \bibinfo {author} {\bibfnamefont {J.~J.~T.}\ \bibnamefont
  {Lindballe}}, \bibinfo {author} {\bibfnamefont {I.~S.}\ \bibnamefont
  {Jensen}}, \bibinfo {author} {\bibfnamefont {P.}~\bibnamefont {Staanum}},
  \bibinfo {author} {\bibfnamefont {D.}~\bibnamefont {Voigt}}, \ and\ \bibinfo
  {author} {\bibfnamefont {M.}~\bibnamefont {Drewsen}},\ }\href {\doibase
  10.1103/PhysRevA.69.042502} {\bibfield  {journal} {\bibinfo  {journal}
  {Physical Review A}\ }\textbf {\bibinfo {volume} {69}},\ \bibinfo {pages}
  {042502} (\bibinfo {year} {2004})}\BibitemShut {NoStop}%
\bibitem [{\citenamefont {Turrin}(1977)}]{Turrin1977}%
  \BibitemOpen
  \bibfield  {author} {\bibinfo {author} {\bibfnamefont {A.}~\bibnamefont
  {Turrin}},\ }\href@noop {} {\bibfield  {journal} {\bibinfo  {journal} {Optics
  Communications}\ }\textbf {\bibinfo {volume} {23}},\ \bibinfo {pages} {220}
  (\bibinfo {year} {1977})}\BibitemShut {NoStop}%
\bibitem [{\citenamefont {Wunderlich}\ \emph {et~al.}(2007)\citenamefont
  {Wunderlich}, \citenamefont {Hannemann}, \citenamefont {K{\"o}rber},
  \citenamefont {H{\"a}ffner}, \citenamefont {Roos}, \citenamefont
  {H{\"a}nsel}, \citenamefont {Blatt},\ and\ \citenamefont
  {Schmidt-Kaler}}]{Wunderlich2007}%
  \BibitemOpen
  \bibfield  {author} {\bibinfo {author} {\bibfnamefont {C.}~\bibnamefont
  {Wunderlich}}, \bibinfo {author} {\bibfnamefont {T.}~\bibnamefont
  {Hannemann}}, \bibinfo {author} {\bibfnamefont {T.}~\bibnamefont
  {K{\"o}rber}}, \bibinfo {author} {\bibfnamefont {H.}~\bibnamefont
  {H{\"a}ffner}}, \bibinfo {author} {\bibfnamefont {C.}~\bibnamefont {Roos}},
  \bibinfo {author} {\bibfnamefont {W.}~\bibnamefont {H{\"a}nsel}}, \bibinfo
  {author} {\bibfnamefont {R.}~\bibnamefont {Blatt}}, \ and\ \bibinfo {author}
  {\bibfnamefont {F.}~\bibnamefont {Schmidt-Kaler}},\ }\href {\doibase
  10.1080/09500340600741082} {\bibfield  {journal} {\bibinfo  {journal}
  {Journal of Modern Optics}\ }\textbf {\bibinfo {volume} {54}},\ \bibinfo
  {pages} {1541} (\bibinfo {year} {2007})}\BibitemShut {NoStop}%
\bibitem [{\citenamefont {Rabi}\ \emph {et~al.}(1939)\citenamefont {Rabi},
  \citenamefont {Millman}, \citenamefont {Kusch},\ and\ \citenamefont
  {Zacharias}}]{Rabi_&al_1939}%
  \BibitemOpen
  \bibfield  {author} {\bibinfo {author} {\bibfnamefont {I.~I.}\ \bibnamefont
  {Rabi}}, \bibinfo {author} {\bibfnamefont {S.}~\bibnamefont {Millman}},
  \bibinfo {author} {\bibfnamefont {P.}~\bibnamefont {Kusch}}, \ and\ \bibinfo
  {author} {\bibfnamefont {J.~R.}\ \bibnamefont {Zacharias}},\ }\href {\doibase
  10.1103/physrev.55.526} {\bibfield  {journal} {\bibinfo  {journal} {Phys.
  Rev.}\ }\textbf {\bibinfo {volume} {55}},\ \bibinfo {pages} {526} (\bibinfo
  {year} {1939})}\BibitemShut {NoStop}%
\bibitem [{sta()}]{stabilaser}%
  \BibitemOpen
  \href@noop {} {}\bibinfo {howpublished}
  {http://www.stabilaser.dk/}\BibitemShut {NoStop}%
\bibitem [{\citenamefont {Talvard}\ \emph {et~al.}(2017)\citenamefont
  {Talvard}, \citenamefont {Westergaard}, \citenamefont {DePalatis},
  \citenamefont {Mortensen}, \citenamefont {Drewsen}, \citenamefont {G{\o}th},\
  and\ \citenamefont {Hald}}]{Talvard2017}%
  \BibitemOpen
  \bibfield  {author} {\bibinfo {author} {\bibfnamefont {T.}~\bibnamefont
  {Talvard}}, \bibinfo {author} {\bibfnamefont {P.~G.}\ \bibnamefont
  {Westergaard}}, \bibinfo {author} {\bibfnamefont {M.~V.}\ \bibnamefont
  {DePalatis}}, \bibinfo {author} {\bibfnamefont {N.~F.}\ \bibnamefont
  {Mortensen}}, \bibinfo {author} {\bibfnamefont {M.}~\bibnamefont {Drewsen}},
  \bibinfo {author} {\bibfnamefont {B.}~\bibnamefont {G{\o}th}}, \ and\
  \bibinfo {author} {\bibfnamefont {J.}~\bibnamefont {Hald}},\ }\href {\doibase
  10.1364/oe.25.002259} {\bibfield  {journal} {\bibinfo  {journal} {Optics
  Express}\ }\textbf {\bibinfo {volume} {25}},\ \bibinfo {pages} {2259}
  (\bibinfo {year} {2017})}\BibitemShut {NoStop}%
\bibitem [{\citenamefont {Johnson}(2007)}]{2007a}%
  \BibitemOpen
  \bibfield  {author} {\bibinfo {author} {\bibfnamefont {W.~R.}\ \bibnamefont
  {Johnson}},\ }\href {\doibase 10.1007/978-3-540-68013-0} {\emph {\bibinfo
  {title} {Atomic Structure Theory}}}\ (\bibinfo  {publisher} {Springer Berlin
  Heidelberg},\ \bibinfo {year} {2007})\BibitemShut {NoStop}%
\bibitem [{\citenamefont {Kurth}\ \emph {et~al.}(1995)\citenamefont {Kurth},
  \citenamefont {Gudjons}, \citenamefont {Hilbert}, \citenamefont {Reisinger},
  \citenamefont {Werth},\ and\ \citenamefont
  {M{\aa}rtensson-Pendrill}}]{Kurth1995}%
  \BibitemOpen
  \bibfield  {author} {\bibinfo {author} {\bibfnamefont {F.}~\bibnamefont
  {Kurth}}, \bibinfo {author} {\bibfnamefont {T.}~\bibnamefont {Gudjons}},
  \bibinfo {author} {\bibfnamefont {B.}~\bibnamefont {Hilbert}}, \bibinfo
  {author} {\bibfnamefont {T.}~\bibnamefont {Reisinger}}, \bibinfo {author}
  {\bibfnamefont {G.}~\bibnamefont {Werth}}, \ and\ \bibinfo {author}
  {\bibfnamefont {A.~M.}\ \bibnamefont {M{\aa}rtensson-Pendrill}},\ }\href
  {\doibase 10.1007/bf01437567} {\bibfield  {journal} {\bibinfo  {journal}
  {Zeitschrift f{\"{u}}r Physik D Atoms, Molecules and Clusters}\ }\textbf
  {\bibinfo {volume} {34}},\ \bibinfo {pages} {227} (\bibinfo {year}
  {1995})}\BibitemShut {NoStop}%
\bibitem [{\citenamefont {Wang}\ \emph {et~al.}(2017)\citenamefont {Wang},
  \citenamefont {Audi}, \citenamefont {Kondev}, \citenamefont {Huang},
  \citenamefont {Naimi},\ and\ \citenamefont {Xu}}]{Wang2017}%
  \BibitemOpen
  \bibfield  {author} {\bibinfo {author} {\bibfnamefont {M.}~\bibnamefont
  {Wang}}, \bibinfo {author} {\bibfnamefont {G.}~\bibnamefont {Audi}}, \bibinfo
  {author} {\bibfnamefont {F.}~\bibnamefont {Kondev}}, \bibinfo {author}
  {\bibfnamefont {W.}~\bibnamefont {Huang}}, \bibinfo {author} {\bibfnamefont
  {S.}~\bibnamefont {Naimi}}, \ and\ \bibinfo {author} {\bibfnamefont
  {X.}~\bibnamefont {Xu}},\ }\href@noop {} {\bibfield  {journal} {\bibinfo
  {journal} {Chinese Physics C}\ }\textbf {\bibinfo {volume} {41}},\ \bibinfo
  {pages} {030003} (\bibinfo {year} {2017})}\BibitemShut {NoStop}%
\bibitem [{\citenamefont {Kramida}\ \emph {et~al.}(2018)\citenamefont
  {Kramida}, \citenamefont {Ralchenko}, \citenamefont {Reader},\ and\
  \citenamefont {{NIST~ASD~Team}}}]{kramida2018}%
  \BibitemOpen
  \bibfield  {author} {\bibinfo {author} {\bibfnamefont {A.}~\bibnamefont
  {Kramida}}, \bibinfo {author} {\bibfnamefont {Y.}~\bibnamefont {Ralchenko}},
  \bibinfo {author} {\bibfnamefont {J.}~\bibnamefont {Reader}}, \ and\ \bibinfo
  {author} {\bibnamefont {{NIST~ASD~Team}}},\ }\href {\doibase
  https://doi.org/10.18434/T4W30F} {\bibfield  {journal} {\bibinfo  {journal}
  {NIST Atomic Spectra Database (version 5.6.1)}\ } (\bibinfo {year} {2018}),\
  https://doi.org/10.18434/T4W30F}\BibitemShut {NoStop}%
\bibitem [{\citenamefont {Boggs}\ \emph {et~al.}(1987)\citenamefont {Boggs},
  \citenamefont {Byrd},\ and\ \citenamefont {Schnabel}}]{Boggs1987}%
  \BibitemOpen
  \bibfield  {author} {\bibinfo {author} {\bibfnamefont {P.~T.}\ \bibnamefont
  {Boggs}}, \bibinfo {author} {\bibfnamefont {R.~H.}\ \bibnamefont {Byrd}}, \
  and\ \bibinfo {author} {\bibfnamefont {R.~B.}\ \bibnamefont {Schnabel}},\
  }\href {\doibase 10.1137/0908085} {\bibfield  {journal} {\bibinfo  {journal}
  {{SIAM} Journal on Scientific and Statistical Computing}\ }\textbf {\bibinfo
  {volume} {8}},\ \bibinfo {pages} {1052} (\bibinfo {year} {1987})}\BibitemShut
  {NoStop}%
\bibitem [{sup()}]{supmat}%
  \BibitemOpen
  \href@noop {} {}\BibitemShut {NoStop}%
\bibitem [{\citenamefont {Flambaum}\ \emph {et~al.}(2018)\citenamefont
  {Flambaum}, \citenamefont {Geddes},\ and\ \citenamefont
  {Viatkina}}]{Flambaum2018}%
  \BibitemOpen
  \bibfield  {author} {\bibinfo {author} {\bibfnamefont {V.~V.}\ \bibnamefont
  {Flambaum}}, \bibinfo {author} {\bibfnamefont {A.~J.}\ \bibnamefont
  {Geddes}}, \ and\ \bibinfo {author} {\bibfnamefont {A.~V.}\ \bibnamefont
  {Viatkina}},\ }\href {\doibase 10.1103/physreva.97.032510} {\bibfield
  {journal} {\bibinfo  {journal} {Physical Review A}\ }\textbf {\bibinfo
  {volume} {97}} (\bibinfo {year} {2018}),\
  10.1103/physreva.97.032510}\BibitemShut {NoStop}%
\bibitem [{\citenamefont {Yao}\ \emph {et~al.}(2006)\citenamefont {Yao} \emph
  {et~al.}}]{Yao:2006px}%
  \BibitemOpen
  \bibfield  {author} {\bibinfo {author} {\bibfnamefont {W.~M.}\ \bibnamefont
  {Yao}} \emph {et~al.} (\bibinfo {collaboration} {Particle Data Group}),\
  }\href {\doibase 10.1088/0954-3899/33/1/001} {\bibfield  {journal} {\bibinfo
  {journal} {J. Phys.}\ }\textbf {\bibinfo {volume} {G33}},\ \bibinfo {pages}
  {1} (\bibinfo {year} {2006})}\BibitemShut {NoStop}%
%%CITATION = JPAGA,G33,1;%%
\bibitem [{\citenamefont {Grifols}\ and\ \citenamefont
  {Masso}(1986)}]{Grifols:1986fc}%
  \BibitemOpen
  \bibfield  {author} {\bibinfo {author} {\bibfnamefont {J.~A.}\ \bibnamefont
  {Grifols}}\ and\ \bibinfo {author} {\bibfnamefont {E.}~\bibnamefont
  {Masso}},\ }\href {\doibase 10.1016/0370-2693(86)90509-5} {\bibfield
  {journal} {\bibinfo  {journal} {Phys. Lett.}\ }\textbf {\bibinfo {volume}
  {B173}},\ \bibinfo {pages} {237} (\bibinfo {year} {1986})}\BibitemShut
  {NoStop}%
%%CITATION = PHLTA,B173,237;%%
\bibitem [{\citenamefont {Grifols}\ \emph {et~al.}(1989)\citenamefont
  {Grifols}, \citenamefont {Masso},\ and\ \citenamefont
  {Peris}}]{Grifols:1988fv}%
  \BibitemOpen
  \bibfield  {author} {\bibinfo {author} {\bibfnamefont {J.~A.}\ \bibnamefont
  {Grifols}}, \bibinfo {author} {\bibfnamefont {E.}~\bibnamefont {Masso}}, \
  and\ \bibinfo {author} {\bibfnamefont {S.}~\bibnamefont {Peris}},\ }\href
  {\doibase 10.1142/S0217732389000381} {\bibfield  {journal} {\bibinfo
  {journal} {Mod. Phys. Lett.}\ }\textbf {\bibinfo {volume} {A4}},\ \bibinfo
  {pages} {311} (\bibinfo {year} {1989})}\BibitemShut {NoStop}%
%%CITATION = MPLAE,A4,311;%%
\bibitem [{\citenamefont {Redondo}\ and\ \citenamefont
  {Raffelt}(2013)}]{Redondo:2013lna}%
  \BibitemOpen
  \bibfield  {author} {\bibinfo {author} {\bibfnamefont {J.}~\bibnamefont
  {Redondo}}\ and\ \bibinfo {author} {\bibfnamefont {G.}~\bibnamefont
  {Raffelt}},\ }\href {\doibase 10.1088/1475-7516/2013/08/034} {\bibfield
  {journal} {\bibinfo  {journal} {JCAP}\ }\textbf {\bibinfo {volume} {1308}},\
  \bibinfo {pages} {034} (\bibinfo {year} {2013})},\ \Eprint
  {http://arxiv.org/abs/1305.2920} {arXiv:1305.2920 [hep-ph]} \BibitemShut
  {NoStop}%
%%CITATION = ARXIV:1305.2920;%%
\bibitem [{\citenamefont {Hardy}\ and\ \citenamefont
  {Lasenby}(2017)}]{Hardy:2016kme}%
  \BibitemOpen
  \bibfield  {author} {\bibinfo {author} {\bibfnamefont {E.}~\bibnamefont
  {Hardy}}\ and\ \bibinfo {author} {\bibfnamefont {R.}~\bibnamefont
  {Lasenby}},\ }\href {\doibase 10.1007/JHEP02(2017)033} {\bibfield  {journal}
  {\bibinfo  {journal} {JHEP}\ }\textbf {\bibinfo {volume} {02}},\ \bibinfo
  {pages} {033} (\bibinfo {year} {2017})},\ \Eprint
  {http://arxiv.org/abs/1611.05852} {arXiv:1611.05852 [hep-ph]} \BibitemShut
  {NoStop}%
%%CITATION = ARXIV:1611.05852;%%
\bibitem [{\citenamefont {Patrignani}\ \emph {et~al.}(2016)\citenamefont
  {Patrignani} \emph {et~al.}}]{Olive:2016xmw}%
  \BibitemOpen
  \bibfield  {author} {\bibinfo {author} {\bibfnamefont {C.}~\bibnamefont
  {Patrignani}} \emph {et~al.} (\bibinfo {collaboration} {Particle Data
  Group}),\ }\href {\doibase 10.1088/1674-1137/40/10/100001} {\bibfield
  {journal} {\bibinfo  {journal} {Chin. Phys.}\ }\textbf {\bibinfo {volume}
  {C40}},\ \bibinfo {pages} {100001} (\bibinfo {year} {2016})}\BibitemShut
  {NoStop}%
%%CITATION = CHPHD,C40,100001;%%
\bibitem [{\citenamefont {Hanneke}\ \emph {et~al.}(2011)\citenamefont
  {Hanneke}, \citenamefont {Hoogerheide},\ and\ \citenamefont
  {Gabrielse}}]{Hanneke:2010au}%
  \BibitemOpen
  \bibfield  {author} {\bibinfo {author} {\bibfnamefont {D.}~\bibnamefont
  {Hanneke}}, \bibinfo {author} {\bibfnamefont {S.~F.}\ \bibnamefont
  {Hoogerheide}}, \ and\ \bibinfo {author} {\bibfnamefont {G.}~\bibnamefont
  {Gabrielse}},\ }\href {\doibase 10.1103/PhysRevA.83.052122} {\bibfield
  {journal} {\bibinfo  {journal} {Physical Review}\ }\textbf {\bibinfo {volume}
  {A83}},\ \bibinfo {pages} {052122} (\bibinfo {year} {2011})},\ \Eprint
  {http://arxiv.org/abs/1009.4831} {arXiv:1009.4831 [physics.atom-ph]}
  \BibitemShut {NoStop}%
%%CITATION = ARXIV:1009.4831;%%
\bibitem [{\citenamefont {Barbieri}\ and\ \citenamefont
  {Ericson}(1975)}]{Barbieri:1975xy}%
  \BibitemOpen
  \bibfield  {author} {\bibinfo {author} {\bibfnamefont {R.}~\bibnamefont
  {Barbieri}}\ and\ \bibinfo {author} {\bibfnamefont {T.~E.~O.}\ \bibnamefont
  {Ericson}},\ }\href {\doibase 10.1016/0370-2693(75)90073-8} {\bibfield
  {journal} {\bibinfo  {journal} {Phys. Lett.}\ }\textbf {\bibinfo {volume}
  {57B}},\ \bibinfo {pages} {270} (\bibinfo {year} {1975})}\BibitemShut
  {NoStop}%
%%CITATION = PHLTA,57B,270;%%
\bibitem [{\citenamefont {Leeb}\ and\ \citenamefont
  {Schmiedmayer}(1992)}]{Leeb:1992qf}%
  \BibitemOpen
  \bibfield  {author} {\bibinfo {author} {\bibfnamefont {H.}~\bibnamefont
  {Leeb}}\ and\ \bibinfo {author} {\bibfnamefont {J.}~\bibnamefont
  {Schmiedmayer}},\ }\href {\doibase 10.1103/PhysRevLett.68.1472} {\bibfield
  {journal} {\bibinfo  {journal} {Physical Review Letters}\ }\textbf {\bibinfo
  {volume} {68}},\ \bibinfo {pages} {1472} (\bibinfo {year}
  {1992})}\BibitemShut {NoStop}%
%%CITATION = PRLTA,68,1472;%%
\bibitem [{\citenamefont {Pokotilovski}(2006)}]{Pokotilovski:2006up}%
  \BibitemOpen
  \bibfield  {author} {\bibinfo {author} {\bibfnamefont {{\relax Yu}.~N.}\
  \bibnamefont {Pokotilovski}},\ }\href {\doibase 10.1134/S1063778806060020}
  {\bibfield  {journal} {\bibinfo  {journal} {Phys. Atom. Nucl.}\ }\textbf
  {\bibinfo {volume} {69}},\ \bibinfo {pages} {924} (\bibinfo {year} {2006})},\
  \Eprint {http://arxiv.org/abs/hep-ph/0601157} {arXiv:hep-ph/0601157 [hep-ph]}
  \BibitemShut {NoStop}%
%%CITATION = HEP-PH/0601157;%%
\bibitem [{\citenamefont {Nesvizhevsky}\ \emph {et~al.}(2008)\citenamefont
  {Nesvizhevsky}, \citenamefont {Pignol},\ and\ \citenamefont
  {Protasov}}]{Nesvizhevsky:2007by}%
  \BibitemOpen
  \bibfield  {author} {\bibinfo {author} {\bibfnamefont {V.~V.}\ \bibnamefont
  {Nesvizhevsky}}, \bibinfo {author} {\bibfnamefont {G.}~\bibnamefont
  {Pignol}}, \ and\ \bibinfo {author} {\bibfnamefont {K.~V.}\ \bibnamefont
  {Protasov}},\ }\href {\doibase 10.1103/PhysRevD.77.034020} {\bibfield
  {journal} {\bibinfo  {journal} {Physical Review}\ }\textbf {\bibinfo {volume}
  {D77}},\ \bibinfo {pages} {034020} (\bibinfo {year} {2008})},\ \Eprint
  {http://arxiv.org/abs/0711.2298} {arXiv:0711.2298 [hep-ph]} \BibitemShut
  {NoStop}%
%%CITATION = ARXIV:0711.2298;%%
\bibitem [{\citenamefont {Bordag}\ \emph {et~al.}(2001)\citenamefont {Bordag},
  \citenamefont {Mohideen},\ and\ \citenamefont
  {Mostepanenko}}]{Bordag:2001qi}%
  \BibitemOpen
  \bibfield  {author} {\bibinfo {author} {\bibfnamefont {M.}~\bibnamefont
  {Bordag}}, \bibinfo {author} {\bibfnamefont {U.}~\bibnamefont {Mohideen}}, \
  and\ \bibinfo {author} {\bibfnamefont {V.~M.}\ \bibnamefont {Mostepanenko}},\
  }\href {\doibase 10.1016/S0370-1573(01)00015-1} {\bibfield  {journal}
  {\bibinfo  {journal} {Phys. Rept.}\ }\textbf {\bibinfo {volume} {353}},\
  \bibinfo {pages} {1} (\bibinfo {year} {2001})},\ \Eprint
  {http://arxiv.org/abs/quant-ph/0106045} {arXiv:quant-ph/0106045 [quant-ph]}
  \BibitemShut {NoStop}%
%%CITATION = QUANT-PH/0106045;%%
\bibitem [{\citenamefont {Bordag}\ \emph {et~al.}(2009)\citenamefont {Bordag},
  \citenamefont {Klimchitskaya}, \citenamefont {Mohideen},\ and\ \citenamefont
  {Mostepanenko}}]{bordag2009advances}%
  \BibitemOpen
  \bibfield  {author} {\bibinfo {author} {\bibfnamefont {M.}~\bibnamefont
  {Bordag}}, \bibinfo {author} {\bibfnamefont {G.~L.}\ \bibnamefont
  {Klimchitskaya}}, \bibinfo {author} {\bibfnamefont {U.}~\bibnamefont
  {Mohideen}}, \ and\ \bibinfo {author} {\bibfnamefont {V.~M.}\ \bibnamefont
  {Mostepanenko}},\ }\href@noop {} {\bibfield  {journal} {\bibinfo  {journal}
  {Int. Ser. Monogr. Phys.}\ }\textbf {\bibinfo {volume} {145}},\ \bibinfo
  {pages} {1} (\bibinfo {year} {2009})}\BibitemShut {NoStop}%
%%CITATION = IMPHA,145,1;%%
\bibitem [{\citenamefont {M{\"u}ller}\ \emph {et~al.}()\citenamefont
  {M{\"u}ller}, \citenamefont {K{\"o}nig}, \citenamefont {Imgram},
  \citenamefont {Kr{\"a}mer},\ and\ \citenamefont
  {N{\"o}rtersh{\"a}user}}]{Muller2020}%
  \BibitemOpen
  \bibfield  {author} {\bibinfo {author} {\bibfnamefont {P.}~\bibnamefont
  {M{\"u}ller}}, \bibinfo {author} {\bibfnamefont {K.}~\bibnamefont
  {K{\"o}nig}}, \bibinfo {author} {\bibfnamefont {P.}~\bibnamefont {Imgram}},
  \bibinfo {author} {\bibfnamefont {J.}~\bibnamefont {Kr{\"a}mer}}, \ and\
  \bibinfo {author} {\bibfnamefont {W.}~\bibnamefont {N{\"o}rtersh{\"a}user}},\
  }\href@noop {} {\bibinfo  {journal} {To be published soon: ``Collinear laser
  spectroscopy of Ca$^+$: Solving the field-shift puzzle of the 4s
  $^2$S$_{1/2}$ - 4p $^2$P$_{1/2, 3/2}$ transitions''}\ }\BibitemShut {NoStop}%
\bibitem [{\citenamefont {Shi}\ \emph {et~al.}(2016)\citenamefont {Shi},
  \citenamefont {Gebert}, \citenamefont {Gorges}, \citenamefont {Kaufmann},
  \citenamefont {N{\"o}rtersh{\"a}user}, \citenamefont {Sahoo}, \citenamefont
  {Surzhykov}, \citenamefont {Yerokhin}, \citenamefont {Berengut},
  \citenamefont {Wolf}, \citenamefont {Heip},\ and\ \citenamefont
  {Schmidt}}]{Shi2016}%
  \BibitemOpen
\bibfield  {journal} {  }\bibfield  {author} {\bibinfo {author} {\bibfnamefont
  {C.}~\bibnamefont {Shi}}, \bibinfo {author} {\bibfnamefont {F.}~\bibnamefont
  {Gebert}}, \bibinfo {author} {\bibfnamefont {C.}~\bibnamefont {Gorges}},
  \bibinfo {author} {\bibfnamefont {S.}~\bibnamefont {Kaufmann}}, \bibinfo
  {author} {\bibfnamefont {W.}~\bibnamefont {N{\"o}rtersh{\"a}user}}, \bibinfo
  {author} {\bibfnamefont {B.~K.}\ \bibnamefont {Sahoo}}, \bibinfo {author}
  {\bibfnamefont {A.}~\bibnamefont {Surzhykov}}, \bibinfo {author}
  {\bibfnamefont {V.~A.}\ \bibnamefont {Yerokhin}}, \bibinfo {author}
  {\bibfnamefont {J.~C.}\ \bibnamefont {Berengut}}, \bibinfo {author}
  {\bibfnamefont {F.}~\bibnamefont {Wolf}}, \bibinfo {author} {\bibfnamefont
  {J.~C.}\ \bibnamefont {Heip}}, \ and\ \bibinfo {author} {\bibfnamefont
  {P.~O.}\ \bibnamefont {Schmidt}},\ }\href {\doibase
  10.1007/s00340-016-6572-z} {\bibfield  {journal} {\bibinfo  {journal}
  {Applied Physics B}\ }\textbf {\bibinfo {volume} {123}},\ \bibinfo {pages}
  {2} (\bibinfo {year} {2016})}\BibitemShut {NoStop}%
\bibitem [{\citenamefont {Kramida}(2020)}]{Kramida2020}%
  \BibitemOpen
  \bibfield  {author} {\bibinfo {author} {\bibfnamefont {A.}~\bibnamefont
  {Kramida}},\ }\href {\doibase 10.1016/j.adt.2019.101322} {\bibfield
  {journal} {\bibinfo  {journal} {Atomic Data and Nuclear Data Tables}\ ,\
  \bibinfo {pages} {101322}} (\bibinfo {year} {2020})}\BibitemShut {NoStop}%
\bibitem [{\citenamefont {Dzuba}\ \emph {et~al.}(1985)\citenamefont {Dzuba},
  \citenamefont {Flambaum}, \citenamefont {Silvestrov},\ and\ \citenamefont
  {Sushkov}}]{dzuba85jpb}%
  \BibitemOpen
  \bibfield  {author} {\bibinfo {author} {\bibfnamefont {V.~A.}\ \bibnamefont
  {Dzuba}}, \bibinfo {author} {\bibfnamefont {V.~V.}\ \bibnamefont {Flambaum}},
  \bibinfo {author} {\bibfnamefont {P.~G.}\ \bibnamefont {Silvestrov}}, \ and\
  \bibinfo {author} {\bibfnamefont {O.~P.}\ \bibnamefont {Sushkov}},\
  }\href@noop {} {\bibfield  {journal} {\bibinfo  {journal} {J. Phys. B}\
  }\textbf {\bibinfo {volume} {18}},\ \bibinfo {pages} {597} (\bibinfo {year}
  {1985})}\BibitemShut {NoStop}%
\bibitem [{\citenamefont {Kahl}\ and\ \citenamefont
  {Berengut}(2019)}]{kahl19cpc}%
  \BibitemOpen
  \bibfield  {author} {\bibinfo {author} {\bibfnamefont {E.~V.}\ \bibnamefont
  {Kahl}}\ and\ \bibinfo {author} {\bibfnamefont {J.~C.}\ \bibnamefont
  {Berengut}},\ }\href@noop {} {\bibfield  {journal} {\bibinfo  {journal}
  {Comp. Phys. Commun.}\ }\textbf {\bibinfo {volume} {238}},\ \bibinfo {pages}
  {232} (\bibinfo {year} {2019})}\BibitemShut {NoStop}%
\bibitem [{\citenamefont {Counts}\ \emph {et~al.}(2020)\citenamefont {Counts},
  \citenamefont {Hur}, \citenamefont {Craik}, \citenamefont {Jeon},
  \citenamefont {Leung}, \citenamefont {Berengut}, \citenamefont {Geddes},
  \citenamefont {Kawasaki}, \citenamefont {Jhe},\ and\ \citenamefont
  {Vuleti\'c}}]{Counts:2020aws}%
  \BibitemOpen
  \bibfield  {author} {\bibinfo {author} {\bibfnamefont {I.}~\bibnamefont
  {Counts}}, \bibinfo {author} {\bibfnamefont {J.}~\bibnamefont {Hur}},
  \bibinfo {author} {\bibfnamefont {D.~P.~A.}\ \bibnamefont {Craik}}, \bibinfo
  {author} {\bibfnamefont {H.}~\bibnamefont {Jeon}}, \bibinfo {author}
  {\bibfnamefont {C.}~\bibnamefont {Leung}}, \bibinfo {author} {\bibfnamefont
  {J.}~\bibnamefont {Berengut}}, \bibinfo {author} {\bibfnamefont
  {A.}~\bibnamefont {Geddes}}, \bibinfo {author} {\bibfnamefont
  {A.}~\bibnamefont {Kawasaki}}, \bibinfo {author} {\bibfnamefont
  {W.}~\bibnamefont {Jhe}}, \ and\ \bibinfo {author} {\bibfnamefont
  {V.}~\bibnamefont {Vuleti\'c}},\ }\href@noop {} {\  (\bibinfo {year}
  {2020})},\ \Eprint {http://arxiv.org/abs/2004.11383} {arXiv:2004.11383
  [physics.atom-ph]} \BibitemShut {NoStop}%
\bibitem [{\citenamefont {Villemoes}\ \emph {et~al.}(1993)\citenamefont
  {Villemoes}, \citenamefont {Arnesen}, \citenamefont {Heijkenskjold},\ and\
  \citenamefont {Wannstrom}}]{Villemoes_1993}%
  \BibitemOpen
  \bibfield  {author} {\bibinfo {author} {\bibfnamefont {P.}~\bibnamefont
  {Villemoes}}, \bibinfo {author} {\bibfnamefont {A.}~\bibnamefont {Arnesen}},
  \bibinfo {author} {\bibfnamefont {F.}~\bibnamefont {Heijkenskjold}}, \ and\
  \bibinfo {author} {\bibfnamefont {A.}~\bibnamefont {Wannstrom}},\ }\href
  {\doibase 10.1088/0953-4075/26/22/030} {\ \textbf {\bibinfo {volume} {26}},\
  \bibinfo {pages} {4289} (\bibinfo {year} {1993})}\BibitemShut {NoStop}%
\bibitem [{\citenamefont {Hucul}\ \emph {et~al.}(2017)\citenamefont {Hucul},
  \citenamefont {Christensen}, \citenamefont {Hudson},\ and\ \citenamefont
  {Campbell}}]{Hucul:2017Ba}%
  \BibitemOpen
  \bibfield  {author} {\bibinfo {author} {\bibfnamefont {D.}~\bibnamefont
  {Hucul}}, \bibinfo {author} {\bibfnamefont {J.~E.}\ \bibnamefont
  {Christensen}}, \bibinfo {author} {\bibfnamefont {E.~R.}\ \bibnamefont
  {Hudson}}, \ and\ \bibinfo {author} {\bibfnamefont {W.~C.}\ \bibnamefont
  {Campbell}},\ }\href {\doibase 10.1103/PhysRevLett.119.100501} {\bibfield
  {journal} {\bibinfo  {journal} {Phys. Rev. Lett.}\ }\textbf {\bibinfo
  {volume} {119}},\ \bibinfo {pages} {100501} (\bibinfo {year}
  {2017})}\BibitemShut {NoStop}%
\end{thebibliography}%

\onecolumngrid

%\appendix
\section{Calculation of the electronic NP coefficients}\label{app:Xi}

Electronic coefficients of the Yukawa potential $X_i$ are calculated using the combination of Brueckner orbitals and random phase approximation~(see~\cite{dzuba85jpb} for details) implemented in the \texttt{AMBiT} code~\cite{kahl19cpc}. Briefly, we start with the Dirac-Fock method to generate core electrons and their potential. Valence and virtual orbitals are constructed by diagonalizing $B$ splines over the Dirac-Fock operator. 
To include the effects of core-valence correlations we use the $B$ spline basis orbitals to create an operator $\hat\Sigma$ such that the second-order correlation correction to the energy of a valence orbital $\left| n \right>$ is $\delta E_n^{(2)} = \langle n | \hat\Sigma | n \rangle$. This operator is then added to the Dirac-Fock operator $\hat h_\mathrm{DF}$ to create Brueckner orbitals defined by
\begin{equation}\label{eq:hDF}
(\hat h_\mathrm{DF} + \hat\Sigma) \psi_n^\mathrm{Br} = E_n^\mathrm{Br} \psi_n^\mathrm{Br}.
\end{equation}

At lowest order, the electronic coefficients $X_i$ are simply the matrix elements $X_i = \langle \psi_i^\mathrm{Br} | V_\mathrm{NP} | \psi_i^\mathrm{Br} \rangle$.
However the potential $V_\mathrm{NP}$ also polarises the core, and the resulting modification of the Dirac-Fock potential is a dominant effect for the valence $d$ orbitals, particularly in the limit of large mediator mass $m_\phi$.
These core polarization effects are taken into account in the random phase approximation, which leads to a modified Dirac Fock potential $\delta V_\mathrm{DF}$ and electronic coefficients
\begin{equation}\label{eq:Xi}
X_i = \langle \psi_i^\mathrm{Br} | V_\mathrm{NP} + \delta V_\mathrm{DF} | \psi_i^\mathrm{Br} \rangle.
\end{equation}

\section{Geometric projection method for 4 isotope pairs}
\label{app:bound}
In this work, we make -- along with \cite{Counts:2020aws} -- for the first time 
use of the New Physics implications of isotope shift measurements at high precision in five isotopes, i.e. four independent isotope pairs. 
%For a fit of four isotope pairs of neutral Yb at MHz precision, see Ref.~\cite{Frugiuele2017}.
While Ref.~\cite{Berengut2018} developed the data-driven method for three isotope pairs and two transitions, and Ref.~\cite{Frugiuele2017} performed a fit to measurements of either more isotopes (four isotope pairs of neutral Yb at MHz precision) or transitions (three transitions in Ca$^+$ without the $A'=46$), here we take the four isotope pairs $AA'$ with $A=40$ and $A'=42,\,44,\,46,\,48$ into account by a geometric projection. 
The vectors of the measured isotope shifts
\begin{equation}
    \overrightarrow{\mu \delta\nu_i} \equiv 
\{\mu^{(1)} \delta\nu_i^{AA'_1},\,
  \mu^{(2)} \delta\nu_i^{AA'_2},\,
  \mu^{(3)} \delta\nu_i^{AA'_3},\,
  \mu^{(4)} \delta\nu_i^{AA'_4}\}\,
\end{equation}
where $\mu^{A'}\equiv \mu^{AA'}$,
are of dimension four. Hence a cross product as in Ref.~\cite{Berengut2018} cannot be applied. Therefore, here we construct the measure of the non-linearity based on scalar products.
We define the $4\times 2$ matrix of isotope shift data as
$D = \left(\overrightarrow{\mu \delta\nu_1},\, \overrightarrow{\mu \delta\nu_2}   \right)$.
Then the projection of the (four-dimensional) modified mass shift vector $\mu=\{1,1,1,1\}$ 
onto the plane spanned by the isotope shift vectors 
$\overrightarrow{\mu \delta\nu_1},\, \overrightarrow{\mu \delta\nu_2}$
is given by
\begin{equation}
    \vec{p} = \left[D \cdot \left(D^T D\right)^{-1} D^T \right]\vec{\mu}\,.
\end{equation}
This allows us to calculate the volume of the parallelepiped spanned by the data vectors and the mass shift direction, which is proportional to the area spanned by the four points in the King plot, as
\begin{equation} \label{eq:V}
    V = |\vec{\mu}-\vec{p}|\,
    \sqrt{\left(\overrightarrow{\mu \delta\nu_1}\right)^2\,
    \left(\overrightarrow{\mu \delta\nu_2}\right)^2
    -\left(\overrightarrow{\mu \delta\nu_1}\cdot \overrightarrow{\mu \delta\nu_2}\right)
    }\,.
\end{equation}
By error propagation of the measurements, we obtain the uncertainty $\sigma_V$ of V. The significance of the non-linearity of a King plot,
${\rm NL} = V/\sigma_V$,
is therefore determined purely from the measured isotope shifts and their uncertainties, independent of theory input. The non-linearity of the Ca$^+$ isotope shifts presented in this work is $V/\sigma_V=1.26$.

%\EF{keep or remove this paragraph?}
%We note, however, that in this case the inclusion of four isotope pairs does not necessarily lead to a stricter bound on NP than three isotope pairs because the subsets of three pairs when omitting 42 or 44 result in a smaller fractional uncertainty $\delta V/V$. This observation is in agreement with the fact that these two data points are slightly further away from the fitted line in Fig.~\ref{fig:king_Dfine_729} than the other two points.

The bound plotted in Fig.~\ref{fig:bounds} %uses all four isotope pairs and 
represents as $y_ey_n + 2\sigma_{y_ey_n}$ the approximate $95\%$CL upper bound on
\begin{equation}
  %  y_ey_n = \frac{V(\overrightarrow{\mu \delta\nu_1},\overrightarrow{\mu \delta\nu_2})}{(X_2 - F_{21} X_1)V(\vec{h},\overrightarrow{\mu \delta\nu_1})}
    y_ey_n \left(m_\phi\right) = \frac{V}{\left[X_2\left(m_\phi\right) - F_{21} X_1\left(m_\phi\right)\right]V_{\vec{h}}}\,, \label{eq:yeynV}
\end{equation}
where $V_{\vec{h}}$ is $V$ with 
$\overrightarrow{\mu \delta\nu_2}$
replaced by the NP direction 
$\vec{h}$ with $h_{AA'}=\gamma_{AA'}/\mu_{AA'}$. For a linear coupling of $\phi$ to the nucleus, $\vec{h}\simeq -A \vec{A'}$\,amu. 
The uncertainty $\sigma_{y_ey_n}$ is obtained by error propagation of $V/V_{\vec{h}}$.
For further details about the data-based non-linearity measure see Ref.~\cite{Berengut2018}.

\section{Estimating near-future sensitivities of Ca$^+$, Ba$^+$ and Yb$^+$}
\label{app:projection}
We estimate bounds from improved isotope-shift measurements in Ca$^+$, Ba$^+$ and Yb$^+$ by two different methods. 
First, with Eq.~(\ref{eq:bestcase}) we use the analytic 'best-case' projection from Ref.~\cite{Berengut2018} that does not require prior isotope shift measurements, but only depends on the absolute uncertainties of the isotope shifts, $\sigma_1, \sigma_2$. It neglects a possible alignment of the field shift with the NP and assumes perfect linearity. Therefore the bound is entirely determined by the uncertainty on $y_ey_n$
\begin{align}
	\label{eq:bestcase}
	[\sigma_{y_ey_n}]_{\text{proj}}	\!\ \sim \! 
	  4\pi \frac{ \sqrt{\sigma^2_2 + \sigma^2_1F_{21}^2}  }{ \left(  X_2  - X_1 F_{21} \right)}
	\frac{ A }{ \Delta A_j^{\rm min} \Delta A_j^{\rm max}   } \, .	
\end{align}
As stated in \cite{Berengut2018}, this projection will always indicate a stronger constraint than a bound from data at the same level of frequency uncertainties $\sigma_i$.

As the second method, in order to allow for a non-linearity of about $1\sigma$ expected in any future measurement, we generate a mock data set with the targeted precision. 
We take S-D$_{5/2}$ as the first transition ($i=1$) from previous data (high precision is not required and at least two isotope pairs are sufficient) 
to predict the second transition ($i=2$) D$_{3/2}$ - D$_{5/2}$,
\begin{align}
    \mu\delta\nu_{2}^{AA'} &= K_{21} + F_{21}  \mu\delta\nu_{2}^{AA'} \pm \delta \cdot \mu^{AA'}\,, \label{eq:fake}
\end{align}
where $\delta$ [Hz] is a possible displacement from the straight line. $K_{21}$ and $F_{21}$ can be taken from available data or from theory. 
The generated data set 
$\{\left[\mu\delta\nu_{1}^{AA'}\right]^{\rm data}, 
\left[\mu\delta\nu_{2}^{AA'}\right]^{\rm mock}\}$ 
and its uncertainty $\{\sigma_1 \mu^{AA'}, \sigma_2 \mu^{AA'}\}$ then yields via Eq.~(\ref{eq:yeynV}) a bound on $y_ey_n$.

In the following paragraphs we provide element-specific information.

\underline{Ca$^+$}
We use $\mu\delta\nu_{S-D_{5/2}}$ and electronic coefficients from this paper.

\underline{Ba$^+$}
From Refs. \cite{Villemoes_1993,Hucul:2017Ba} we calculate the isotope shifts
$\delta\nu^{138,A}_{S-D_{5/2}}$, $\delta\nu^{138,A}_{D_{3/2}-D_{5/2}}$ for $A=134, 136$
and obtain
$F_{21}^{\ba} = -0.00491734$
and $K_{21} = -17.454\ghz\amu$ 
to calculate the approximated isotope shifts of the pairs with $A=130, 132$ assuming perfect linearity. The Ba masses are taken from \cite{Wang2017}.
In the evaluation of the bound on $y_ey_n$ according to Eq.~(\ref{eq:yeynV}), we use the theory value $F_{21}^\ba=-0.0141905$ from our calculation with Eq.~(\ref{eq:Xi}) in order to ensure convergence of $X_2/X_1 \rightarrow F_{21}$ in the limit of large $m_\phi$.

\underline{Yb$^+$}
We use $\delta\nu^{AA'}_{S-D_{5/2}}$ from the recent Counts \textit{et al.} \cite{Counts:2020aws} with $A'=A+2$ for the four pairs of $A=168, 170, 172, 174$. Hence the elements of the NP vector $\vec{h}$ are given by
$h_n= - A_n A_{n+1} \amu = - A_n (A_n+2)\amu$.
From \cite{Counts:2020aws} we calculate 
$K_{21}=-65.8224 \ghz\amu$.

As two benchmarks we choose $\sigma_2\equiv\sigma_{DD}=20\hz$ (the current level achieved for Ca$^+$ in this work) 
and 10 mHz (achieved already for Sr$^+$ \cite{Manovitz2019}) and set
$\sigma_{SD} < \sigma_{DD}/F_{21}$ such that the bound will be determined by $\sigma_{DD}$ whereas an improvement of $\sigma_{SD}$ would not improve the bound significantly.
We consider two options, see Tab.~\ref{tab:projections}: 
\textit{(i)} all points are on the straight line (i.e. the non-linearity vanishes by construction);
\textit{(ii)} we choose the displacement $\delta$ such that the significance of the resulting non-linearity is $V/\sigma_V =1$ where $V$ is the volume defined in Eq.~\ref{eq:V} and $\sigma_V$ its uncertainty.

\begin{table}[h]
\begin{center}
\caption{Upper limits $y_ey_n+2\sigma_{y_ey_n}$ expected in the massless limit ($m_\phi=1\ev/c^2$) in two benchmarks. Comparison of limit from mock data from Eq.~(\ref{eq:fake}) with 
$V/\sigma_V \simeq 1\sigma$ nonlinearity, exact linearity ($\delta=0$) and from the simplified estimate of Eq.~(\ref{eq:bestcase}).
Ca$^+$ with isotope pairs $\{42,44,46,48\}-40$,
Ba$^+$ with $\{130,132,134,136\}-138$ and
Yb$^+$ with neighboring pairs of $\{168,170,172,174,176\}$. The last lines contains the field shift ratio from theory and the sensitivity factor $X_2 - F_{21} X_1$ for $m_\phi=1\ev/c^2$. The example of Ca$^+$ with $\sigma_2=20\,\hz$ is for comparison only, instead the bound from real data is plotted.
}
\setlength{\tabcolsep}{20pt}
\renewcommand{\arraystretch}{1.2}
\begin{tabular}{l|lll}
\hline\hline
& Ca$^+$ &Ba$^+$ &Yb$^+$\\ \hline
$\delta,~ \sigma_1,~ \sigma_2$ [Hz] &12, 2000, 20 &12, 100, 20 & 10, 100, 20\\
$V/\sigma_V$ &1.002 &0.999 &0.992 \\
$y_ey_n/\hbar c$ &$5.4\cdot10^{-11}$ &$4.8\cdot 10^{-12}$ &$3.2\cdot 10^{-12}$ \\
$y_ey_n/\hbar c$, $\delta=0$ & $4.3\cdot 10^{-12}$ &$1.74\cdot 10^{-12}$ &$3.9\cdot 10^{-13}$ \\
$y_ey_n/\hbar c$, Eq.~(\ref{eq:bestcase}) &$3.8\cdot 10^{-12}$ &$1.72\cdot 10^{-12}$ &$8.05\cdot 10^{-13}$ \\ \hline
$\delta,~ \sigma_1,~ \sigma_2$ [Hz]&$0.006, 1, 0.01$ &$0.006, 0.1, 0.01$ &0.0055, 0.1, 0.01 \\
$V/\sigma_V$ &1.003 &1.089 &1.07 \\
$y_ey_n/\hbar c$ &$2.7\cdot 10^{-14}$ &$2.4\cdot 10^{-15}$ &$1.7\cdot10^{-15}$ \\
$y_ey_n/\hbar c$, $\delta=0$ &$2.2\cdot10^{-15}$ &$8.68\cdot10^{-16}$ &$2.0\cdot10^{-16}$ \\
$y_ey_n/\hbar c$, Eq.~(\ref{eq:bestcase}) &$1.9\cdot10^{-15}$ &$8.66\cdot 10^{-16}$ &$4.1\cdot10^{-16}$ \\\hline
$F_{21}=F_2/F_1$ &-0.001795 &-0.01419 &-0.02152 \\
$X_2 - F_{21} X1$ [Hz] &$3.353\cdot 10^{14}$ &$-2.527\cdot 10^{15}$ &$-6.908\cdot 10^{15}$\\
\hline\hline
\end{tabular}
\label{tab:projections}
\end{center}
\end{table}

Tab.~\ref{tab:projections} shows that the mock data method with vanishing non-linearity reproduces the limit obtained from the analytic estimate where all points are assumed to be on a perfect line (apart from a difference in Yb$^+$). Furthermore it shows that a non-linearity of $1\sigma$ weakens the bound by up to an order of magnitude. In Fig.~\ref{fig:bounds} we plot the bounds corresponding to a $1\sigma$ non-linearity as a more realistic estimate.

For Ca$^+$, we note that the mock data method with $\sigma_{SD}=2\khz$, $\sigma_{DD}=27\hz$ and $\delta=19\hz$ ($V/\sigma_V=1.2$) yields $y_ey_n/\hbar c<7.6\cdot 10^{-11}$, i.e. it reproduces the bound from real data accurately.
Furthermore, we compare the bound based on $\sigma_{DD}=27\hz$ to the bound from measuring S-D$_{3/2}$ and S-D$_{3/2}$ both at 20\,Hz precision, obtaining for the example of zero non-linearity and $m_\phi=1\ev/c^2$,
$\{5.3,~5.1\}\cdot 10^{-12}$, respectively,
hence confirming their equivalence.

\end{document}